\documentclass[aps,prb,10pt,twocolumn,floatfix,showpacs,superscriptaddress,longbibliography]{revtex4-1}

\usepackage{graphicx,amsmath,amsfonts,bm}
\usepackage[colorlinks=true, citecolor=blue, linkcolor=blue, urlcolor=blue]{hyperref}
\usepackage[all]{hypcap}
\usepackage{bbold}
\usepackage{ulem}

\usepackage{soul}
\usepackage{xcolor}

\begin{document}

\title{Formation and detection of Majorana modes in quantum spin Hall trenches}
\author{C. Fleckenstein}
\email{christoph.fleckenstein@physik.uni-wuerzburg.de}
\affiliation{Institute of Theoretical Physics and Astrophysics, University of W\"urzburg, 97074 W\"urzburg, Germany}
\affiliation{Department of Physics, KTH Royal Institute of Technology, Stockholm, 106 91 Sweden}
\author{N. Traverso Ziani}
\affiliation{Dipartimento di Fisica, Universit\`a di Genova, 16146 Genova, Italy}
\affiliation{CNR spin, 16146 Genova, Italy}
\author{A. Calzona}
\affiliation{Institute of Theoretical Physics and Astrophysics, University of W\"urzburg, 97074 W\"urzburg, Germany}
\author{M. Sassetti}
\affiliation{Dipartimento di Fisica, Universit\`a di Genova, 16146 Genova, Italy}
\affiliation{CNR spin, 16146 Genova, Italy}
\author{B. Trauzettel}
\affiliation{Institute of Theoretical Physics and Astrophysics, University of W\"urzburg, 97074 W\"urzburg, Germany}
\affiliation{W\"urzburg-Dresden Cluster of Excellence ct.qmat, Germany}
\begin{abstract}
We propose a novel realization for a topologically superconducting phase hosting Majorana zero-modes on the basis of quantum spin Hall systems. Remarkably, our proposal is completely free of ferromagnets. Instead, we confine helical edge states around a narrow defect line of finite length in a two-dimensional topological insulator. We demonstrate the formation of a new topological regime, hosting protected Majorana modes in the presence of s-wave superconductivity and Zeeman coupling. Interestingly, when the system is weakly tunnel-coupled to helical edge state reservoirs, a particular transport signature is associated with the presence of a non-Abelian Majorana zero-mode.
\end{abstract}

\maketitle

\section{Introduction}
The theoretical prediction\cite{Kane2005,KaneMele2005b,bernevig2006} and experimental realization\cite{konig2007} of two-dimensional topological insulators marked the beginning of immense research activities in view of their functionalities in spintronics\cite{Nadj-Perge602,Roth294,Trauzettel2013,Linder2015,Breunig2018} and topological quantum computation\cite{Mong2014}. In particular, the formation and detection of topological superconductivity on the basis of topological systems attracted a lot of attention \cite{FuKane2008, FuKane2009, RRDu2012,Hart2014, Finocchiaro2018, Recher2019} and the emergence of topologically protected Majorana bound states came to the forefront of research \cite{Kitaev2001}. The interest in those excitations is both fundamental and practical, since they obey non-Abelian statistics \cite{Ivanov2001,Nayak2008,Alicea2011} and, hence, can potentially be used for topological quantum computation. Regarding the realization of topologically confined Majoranas using topological insulators, the possibility of inducing superconducting pairing \cite{Bocquillon2018} is promising. However, most proposals rely on the coexistence of ferromagnetic ordering \cite{FuKane2009,Akhmerov2009,Tanaka2009,Crepin2014}, which turns out to be difficult to achieve in the laboratory.

In parallel, another platform for topological superconductivity was found by the prediction of Majorana zero-modes in spin-orbit coupled quantum wires\cite{vonOppen2010,Lutchyn2010}. Subsequently, several experimental works were able to confirm some of the proposed signatures \cite{Kouwenhoven2012,Marcus2016,Albrecht2016}. However, the ultimate proof of the existence of Majoranas is probably still missing.

\begin{figure}
\centering
\includegraphics[scale=0.6]{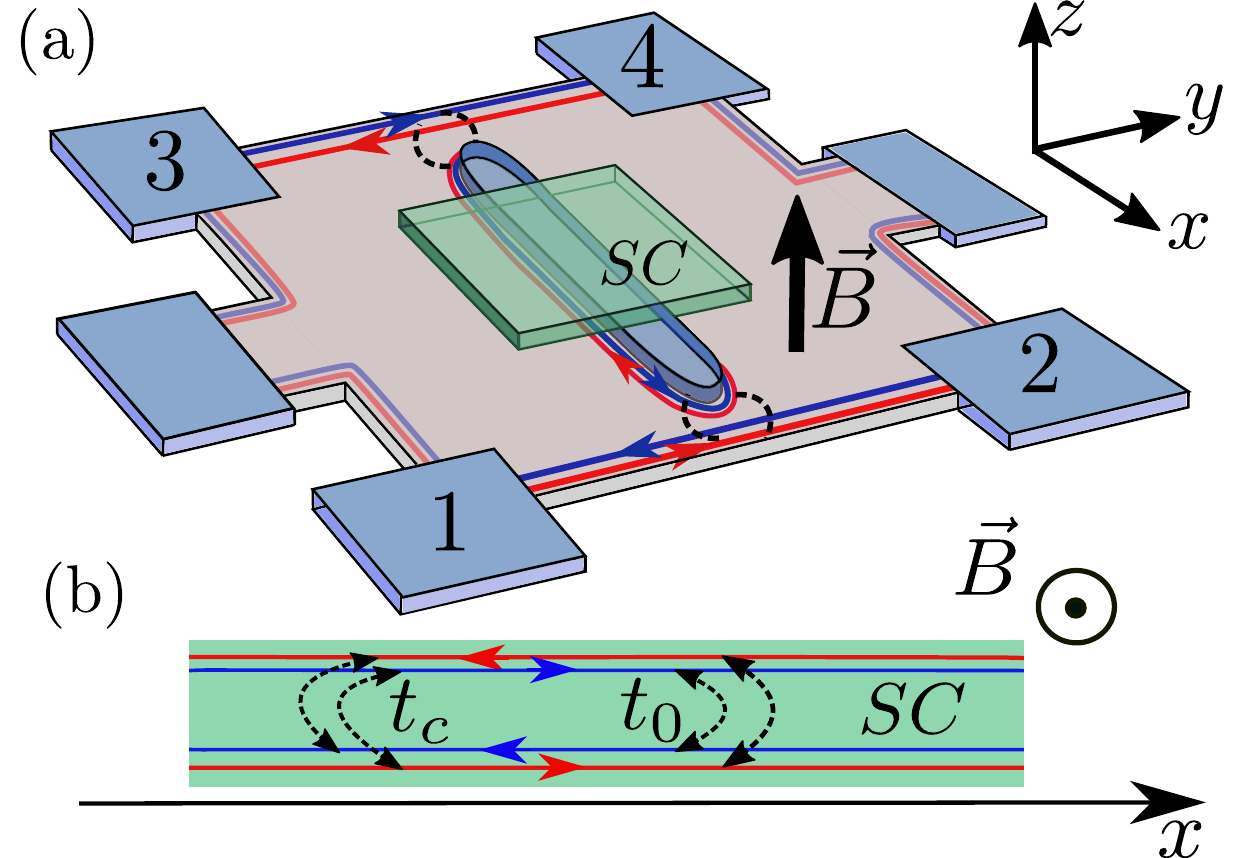}
\caption{\textbf{Quantum spin Hall anti-wire}. (a) Schematic illustration of the system: A QSH anti-wire, covered by a s-wave superconductor under the influence of a magnetic field weakly coupled to helical edge states at the boundary of the QSH stripe. (b) Sketch of the QSH constriction with the appearing scattering terms.}
\label{Fig:1}
\end{figure}

In this work, we propose a hybrid structure that combines the features of topological edge states and spin-orbit coupled quantum wires. The system we investigate -- a quantum spin Hall (QSH) anti-wire -- defines itself through a narrow slit in a two-dimensional topological insulator (see Fig.~\ref{Fig:1}). This system shares similarities with QSH quantum point contacts, recently realized in the laboratory \cite{Strunz2020}, for which the formation of Kramers pairs of Majorana fermions and other complex anyons were proposed \cite{KlinovajaLossQPC2014,LossParaOrtho,Lutchyn2016,Fleckenstein2018,Fleckenstein2019Floquet}. We demonstrate below that the QSH anti-wire, in the presence of s-wave pairing and Zeeman coupling, possesses a topological phase hosting Majorana end-modes. This phase emerges if the slit is narrow enough such that the edge states at opposite sides overlap.
This setup offers key advantages with respect to other platforms. Indeed, the emergence of Majorana modes within a two-dimensional topological insulator makes it straightforward to couple them to topological edge channels, whose helical nature allows for richer transport signatures than a standard tunneling probe. In particular, in the multi-terminal conductance $G_{1\rightarrow 2}=\mathrm{d}I_2/\mathrm{d}V_1$, between contacts $1$ and $2$ of Fig.~\ref{Fig:1} (a), we identify a qualitative Majorana signature beyond the well-known zero-bias peak: The presence of a Majorana-like state at zero energy gives rise to a negative $G_{1\rightarrow 2}$, which is otherwise positive. 
In addition to that, our setup can be easily scaled-up by carving several slits within the same topological insulator. The resulting collection of localized Majorana modes, which can be manipulated by tuning their pair-wise couplings via top gates, would represent a convenient playground for topological quantum computation applications.  

The article is organized as follows. In Sec. \ref{sec:model_results} we discuss the topological properties of narrow QSH trenches. Subsequently, in Sec. \ref{sec:majorana}, we investigate the formation of topologically protected Majorana modes associated with the topological phase. This is followed by a discussion of possible transport signatures in Secs. \ref{sec:transport}, \ref{sec:toy_model} and \ref{sec:robustness}. Finally, we conclude in Sec. \ref{sec:discussion}, where we summarize the results.

\section{Topological phase transition in the anti-wire}
\label{sec:model_results}
\begin{figure}
\centering
\includegraphics[scale=0.2]{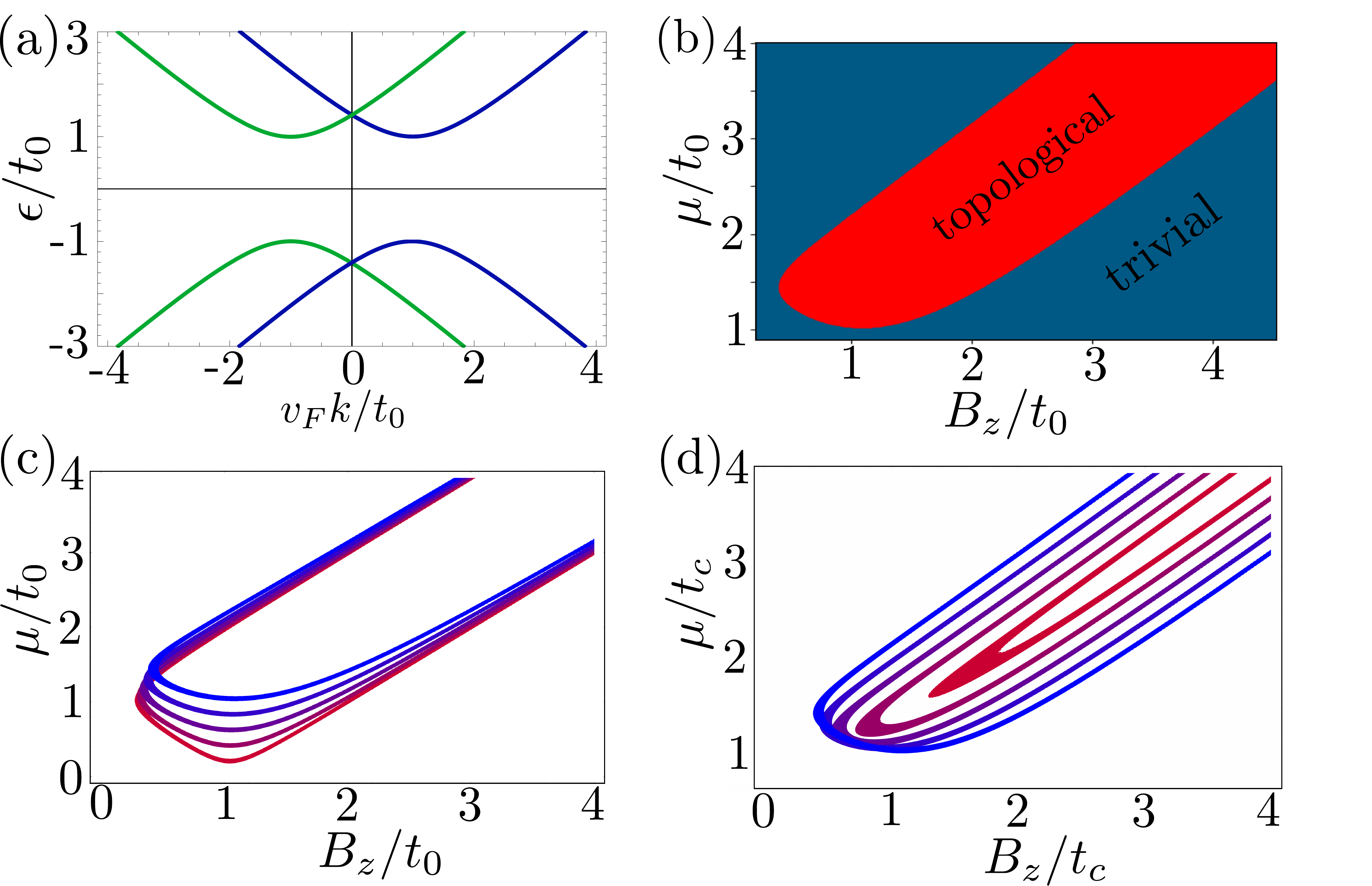}
\label{Fig:spectrum}
\caption{\textbf{Topological phase diagram of the proximitized anti-wire.} (a) Eigenenergy spectrum of $H_0$. The different colors represent states with orthogonal spin with $t_c=t_0$. (b) Phase diagram as function of $\mu$ and $B_z$ (under the choice $t_0=t_c=1$, $\Delta/t_0=0.3$,  $v_F=1$). (c) Dependence of the topological phase on $t_c$. The different curves correspond to gap closures for $t_c/t_0=0.2,0.4,0.6,0.8,1.0$ (red to blue), $\Delta=0.3 t_0,~t_0=1$, $v_F=1$. (d) Dependence of the topological phase on $t_0$. The curves correspond to to gap closures for $t_0/t_c=0.2,0.4,0.6,0.8,1$ (red to blue), $\Delta/t_c=0.3,~t_c=1$, $v_F=1$.}
\end{figure}

The setup we propose is sketched in Fig.~\ref{Fig:1} (a). Its innovative ingredient is a long quantum constriction between two metallic edges of a quantum spin Hall insulator depicted in Fig.~\ref{Fig:1} (b). To compute its topological properties, we first consider the limit of an infinitely long constriction. The kinetic energy can be described by the effective Hamiltonian density $(\hbar = 1)$
\begin{equation}
\label{Eq:Hp}
\mathcal{H}_p = \sum_{\nu,\sigma}\hat{\psi}_{\nu,\sigma}^{\dagger}(x) (-iv_F\sigma\nu\partial_x-\mu) \hat{\psi}_{\nu,\sigma}(x),
\end{equation}
where $\hat{\psi}_{\nu,\sigma}(x)$ are annihilating fermionic fields carrying spin-index $\sigma \in \{\uparrow,\downarrow\}=\{+,-\}$ and edge-index $\nu \in \{1,2\}=\{+,-\}$; $\mu$ acts as a chemical potential and $v_F$ is the Fermi velocity (estimated to be $(10^{5}-10^{6})\mathrm{m/s}$ for QSH systems based on Hg(Cd)Te quantum wells \cite{Konig2008}). We assume a finite overlap of wave functions from states at different sides of the anti-wire. In presence of time-reversal (TR) symmetry, two single particle terms emerge \cite{Teo2009,Liu2011,Lutchyn2016,Dolcini2011, Schmidt2012}
\begin{eqnarray}
\label{Eq:Ht0}
\mathcal{H}_{t_0}&=&t_0 \sum_{\sigma}\left[\hat{\psi}^{\dagger}_{1,\sigma}(x)\hat{\psi}_{2,\sigma}(x)+\mathrm{h.c.}\right],\\
\label{Eq:Htc}
\mathcal{H}_{t_c}&=&t_c  \sum_{\nu}\left[\nu\hat{\psi}_{\nu,\uparrow}^{\dagger}(x)\hat{\psi}_{-\nu,\downarrow}(x)+\mathrm{h.c.}\right].
\end{eqnarray}
While Eq.~(\ref{Eq:Ht0}) describes a hybridization of fermionic states with the same spin associated to different sides of the slit and does not require further symmetry breaking with respect to $\mathcal{H}_p$, Eq.~(\ref{Eq:Htc}) is only finite if axial spin symmetry is absent and takes the role of an effective spin-orbit coupling across the slit \cite{Wu2006}. The spectrum associated with $H_0=\int_{-\infty}^{+\infty} dx\, [\mathcal{H}_p+\mathcal{H}_{t_0}+\mathcal{H}_{t_c}]$ is shown in Fig.~\ref{Fig:spectrum} (a). The additional application of a Zeeman field perpendicular to the $x$ direction opens a partial gap around $k=0$. For concreteness, we consider a field along the $z$ direction
\begin{equation}
\label{Eq:HB}
\mathcal{H}_B =B_z\sum_{\nu,\sigma}\sigma\hat{\psi}_{\nu,\sigma}^{\dagger}(x)\hat{\psi}_{\nu,\sigma}(x).
\end{equation}
The gyro-magnetic factor for the edge states is predicted to be $g\sim 10$ \cite{Wimmer2018} for typical QSH materials. Moreover, the typical values for the effective electron mass in HgTe quantum wells \cite{e_mass} indicate that indeed a situation similar to hybrid systems based on spin-orbit nanowires is met \cite{wire_values}. This implies required magnetic fields of the order of few $\mathrm{mT}$, compatible with the presence of superconductivity.

The resulting band structure shares similarities with spin-orbit  nanowires under the influence of magnetic fields. It can hence be expected that topological physics emerges when s-wave superconductivity is taken into account via
\begin{equation}
\label{Eq:HD}
\mathcal{H}_{\Delta}=\Delta\sum_{\nu}\left[\hat{\psi}_{\nu,\uparrow}^{\dagger}(x)\hat{\psi}_{\nu,\downarrow}^{\dagger}(x)+\mathrm{h.c.}\right].
\end{equation}
Typical values for the proximity induced superconducting order parameter $\Delta$ are given by $\Delta \sim 40 \mathrm{\mu eV}$ in HgTe-based systems\cite{Bocquillon2018}.
Indeed, the infinitely long anti-wire described by $H_0+\int_{-\infty}^{+\infty}dx\,  [\mathcal{H}_{\Delta}+\mathcal{H}_B]$ undergoes a topological phase transition, indicated by a gap-closing and reopening depending on the control parameters $\mu$ and $B_z$ (see Fig.~\ref{Fig:spectrum} (b)). Since the coupling strength $t_c$ in Eq.~(\ref{Eq:Htc}) effectively takes the role of a spin-orbit coupling, as long as it is non-zero, it hardly affects the topological parameter regime [see Fig.~\ref{Fig:spectrum} (c)]. However, it  controls the magnitude of the gaps in the topological regime and therefore the decay length of possible low-energy bound states in the presence of boundaries. By contrast, Eq.~(\ref{Eq:Ht0}) has less influence on the magnitude of the gaps, but strongly affects the shape of the topological regime (Fig.~\ref{Fig:spectrum} (d)). While a concrete estimation of the magnitude of $t_0$ is difficult, it is clear that it can be tuned, up to the magnitude of the bulk gap, by reducing the width of the slit \cite{Moore2010}.

\section{Topologically protected Majoranas}
\label{sec:majorana}
To investigate the presence of topological bound states, we now focus on a slit with a finite length $L$. It is convenient to consider the additional Hamiltonian density
\begin{equation}
\label{Eq:HT}
\mathcal{H}_{T}\!= T\left[\delta(x)\!+\!\delta(x\!-\!L)\right]\sum_{\sigma}\left[\hat{\psi}^{\dagger}_{1,\sigma}(x)\hat{\psi}_{2,\sigma}(x)\!+\!\mathrm{h.c.}\right],
\end{equation}
which describes the presence of barriers at $x=0$ and $x=L$. Indeed, in the limit $T\to\infty$, the Hamiltonian $H_{\textrm AW} =  \lim_{T\to\infty}  \int_0^L dx\, [ \mathcal{H}_p+\mathcal{H}_{t_0}+\mathcal{H}_{t_c} +\mathcal{H}_B + \mathcal{H}_\Delta + \mathcal{H}_T]$
defines an isolated antiwire in the region $x\in[0,L]$, whose fermionic fields obey the open boundary conditions (BCs) (see also App. \ref{sec:app_BC}) \cite{Dolcetto2013}
\begin{equation}
\label{Eq:BC5}
\begin{array}{lcr}
\hat{\psi}_{1,\uparrow}(x) = i \hat{\psi}_{2,\uparrow}(-x), \\
\hat{\psi}_{2,\downarrow}(x) = i \hat{\psi}_{1,\downarrow}(-x),
\end{array}
\end{equation}
where $\hat{\psi}_{\nu,\sigma}(x)=\sum_q \psi_{\nu,\sigma,q}(x)\hat{c}_q$ with annihilation operators $\hat{c}_q$ and the quantization condition $q = (\pi/L)(n-1/2)$.
\begin{figure}
\centering
\includegraphics[scale=0.3]{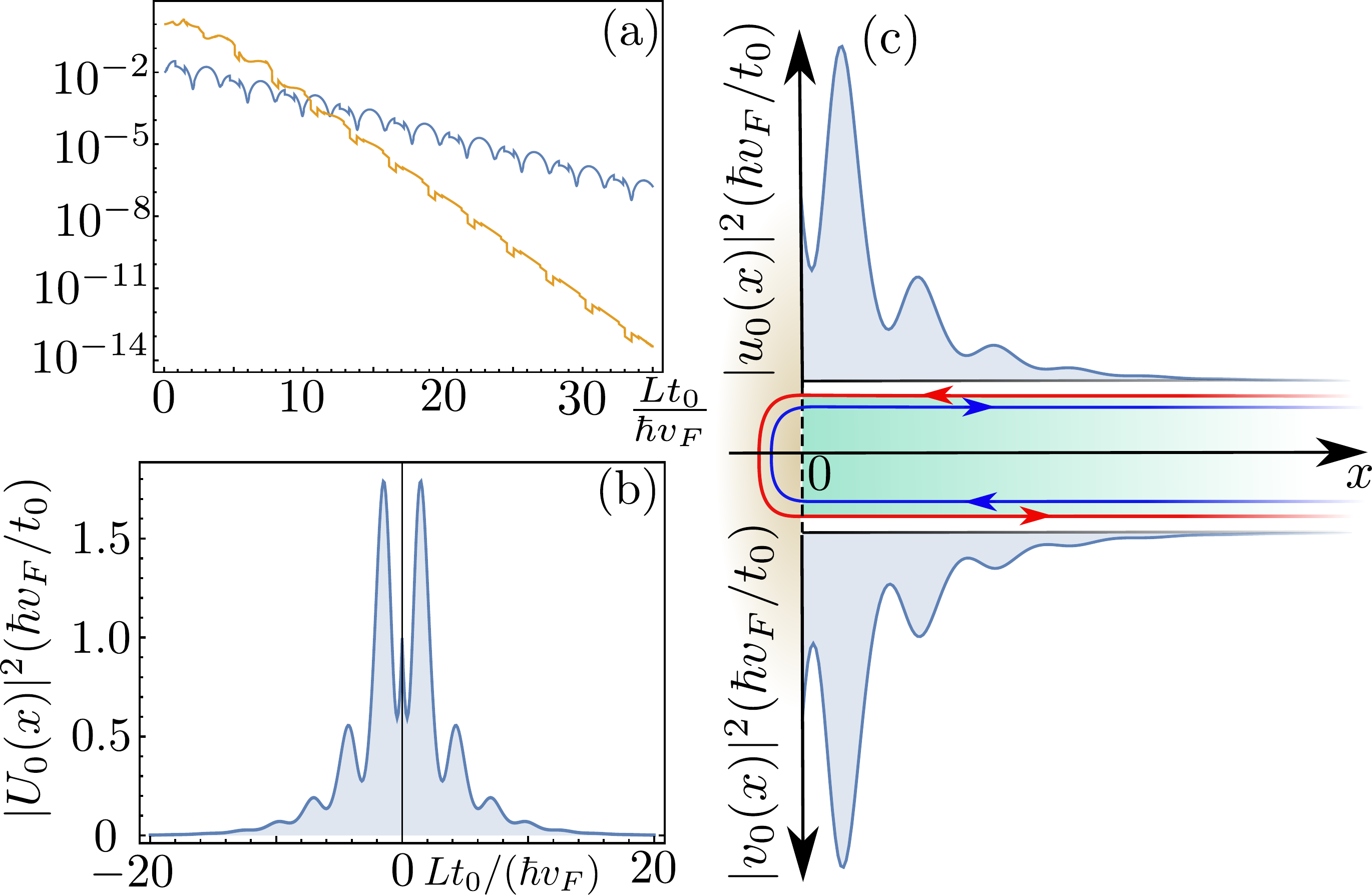}
\label{Fig:wave}
\caption{\textbf{Majorana wavefunctions at the anti-wire ends.} (a) $\lambda_M$ (yellow) and $\delta \Gamma_{\lambda_M}$ (blue) as a function of $L$. (b)  $\vert U_{0}(x)\vert^2$ according to Eq.~(\ref{Eq:BC8}) with $U_{0}(0)=\nu_{\lambda}$. (c) Schematic illustration of the probability distribution in the (folded) anti-wire. The parameters of the calculation are: $B/t_0=0.6$, $\mu/t_0 = \sqrt{2}$, $\Delta/t_0=0.3$, $t_c=t_0=1$, $v_F=1$.}
\end{figure}
We hence obtain
\begin{eqnarray}
\label{Eq:BC6}
H_{\mathrm{AW}}\! &=& \!\int_{-L}^L\!\mathrm{dx}\, \hat{\Phi}^{\dagger}(x)\big[-iv_F\partial_x+\tau_z\sigma_z B_z+\tau_z\sigma_0\mu\nonumber\\
&+&\tau_z\sigma_x~\mathrm{sign}(x)t_c\big]\hat{\Phi}(x)\nonumber\\
&-&\!\int_{-L}^L\!\mathrm{dx}\, \Phi^{\dagger}(x)\big[\tau_x\sigma_y \Delta+i~\mathrm{sign}(x)t_0\big]\hat{\Phi}(-x),
\end{eqnarray}
where $\tau_j$, $\sigma_j$ ($j \in \{x,y,z\}$) are Pauli matrices acting on particle-hole, spin-space, respectively, and $\hat{\Phi}(x)= (\hat{\psi}_{1,\uparrow}(x),\hat{\psi}_{2,\downarrow}(x),\hat{\psi}_{1,\uparrow}^{\dagger}(x),\hat{\psi}_{2,\downarrow}^{\dagger}(x))^T$. Our goal is to determine the eigenfunctions $U_{\epsilon}(x)$ of the Hamiltonian density in Eq.~(\ref{Eq:BC6}). We can overcome its non-locality with the ansatz
\begin{eqnarray}
\label{Eq:BC8}
U_{\epsilon}(x)=u_{\epsilon}(x)\theta(x)+v_{\epsilon}(-x)\theta(-x),
\end{eqnarray}
where $u_{\epsilon}(x)$ and $v_{\epsilon}(x)$ are spinors in the given basis. From the continuity of the solutions $U_{\epsilon}(x)$ at $x=0$ as well as from the anti-periodicity of the system with respect to $2L$, the solution needs to obey the BCs $u_{\epsilon}(0)=v_{\epsilon}(0)$ and $u_{\epsilon}(L)=-v_{\epsilon}(L)$. The single particle problem associated with Eq.~(\ref{Eq:BC6}) becomes equivalent to the set of equations for the functions $u_{\epsilon}(x)$ and  $v_{\epsilon}(x)$ and the eigenenergies $\epsilon$
\begin{eqnarray}
\label{Eq:BC10}
\big[&-&iv_F\partial_{x}s_z\tau_0\sigma_0+s_0\tau_z\sigma_z B_z+s_0\tau_z\sigma_0 \mu+ s_z\tau_z\sigma_x t_c\nonumber\\
&-&s_x\tau_x\sigma_y \Delta + s_y\tau_0\sigma_0 t_0\big] \chi_{\epsilon}(x) = \epsilon\chi_{\epsilon}(x),
\end{eqnarray}
where we define the basis function $\chi_{\epsilon}(x)=(u_{\epsilon}(x),v_{\epsilon}(x))^T$ and the Pauli matrices $s_j$ acting on the space spanned by $u_{\epsilon}(x)$ and $v_{\epsilon}(x)$. The general solution of Eq.~(\ref{Eq:BC10}) can be found by integration
\begin{eqnarray}
\label{Eq:BC11}
\chi_{\epsilon}(x)&=&M_{\epsilon}(x,x_0)\chi_{\epsilon}(x_0),
\end{eqnarray}
where
\begin{equation}
	\begin{split}
	&M_{\epsilon}(x,x_0)=\exp\!\bigg[\int_{x_0}^{x} \mathrm{dx'}\frac{i}{v_F}s_z\tau_0\sigma_0\big(\epsilon-(s_0\tau_z\sigma_z B_z+\\&\quad 
	+s_0\tau_z\sigma_0 \mu + s_z\tau_z\sigma_x t_c-s_x\tau_x\sigma_y \Delta + s_y\tau_0\sigma_0 t_0)\big)\!\bigg].
	\end{split}
\end{equation}
Not every energy $\epsilon$ is compatible with the BCs. For the topological phase, however, in the limit $L\rightarrow \infty$ there should be a decaying solution for $\epsilon \rightarrow 0$ of the form $\Gamma(0)=(\zeta(0),\zeta(0))$ (fulfilling the BCs at $x=0$). Thus, in this limit, Eq.~(\ref{Eq:BC11}) turns into an eigenvalue problem for $\zeta(0)$ of the form
\begin{eqnarray}
\label{Eq:BC12}
\lim_{L\rightarrow \infty} M_0(L,0)\Gamma(0) \stackrel{!}{=} 0.
\end{eqnarray}
If we further demand the solution to be a Majorana, we require $\zeta(0)=\big(f(0),g(0),f^*(0),g^*(0)\big)^T$. Note that demanding a Majorana from of $\Gamma(0)$ implies this form to remain for any other point $x$ because of the particle-hole symmetry of $M_0(x,x')$.
For finite $L$, Eq.~(\ref{Eq:BC12}) does not hold anymore. However, we find that an approximate Majorana solution exists, i.e. $M_0(L,0)$ possesses an eigenvalue $\lambda_M \sim \exp(-\alpha L)$ whose corresponding eigenvector $\nu_{\lambda_M}$ fulfils the BC at $x=0$ and deviates by $\delta\Gamma_{\lambda_M} = \frac{1}{2}\big\vert\!\big\vert s_0(\mathbb{1}-\tau_x\sigma_0)\mathrm{Re}[\nu_{\lambda_M}]\!+\!s_0(\mathbb{1}+\tau_x\sigma_0)\mathrm{Im}[\nu_{\lambda_M}]\big\vert\!\big\vert \sim \exp(-\beta L)$ ($\alpha, \beta \in \mathbb{R}$) from the Majorana form (see Fig.~\ref{Fig:wave} (a)). The probability density associated to the wavefunction is shown in Fig.\ref{Fig:wave} (b,c).

\begin{figure*}[t]
\centering

\includegraphics[width=\textwidth,scale=0.25]{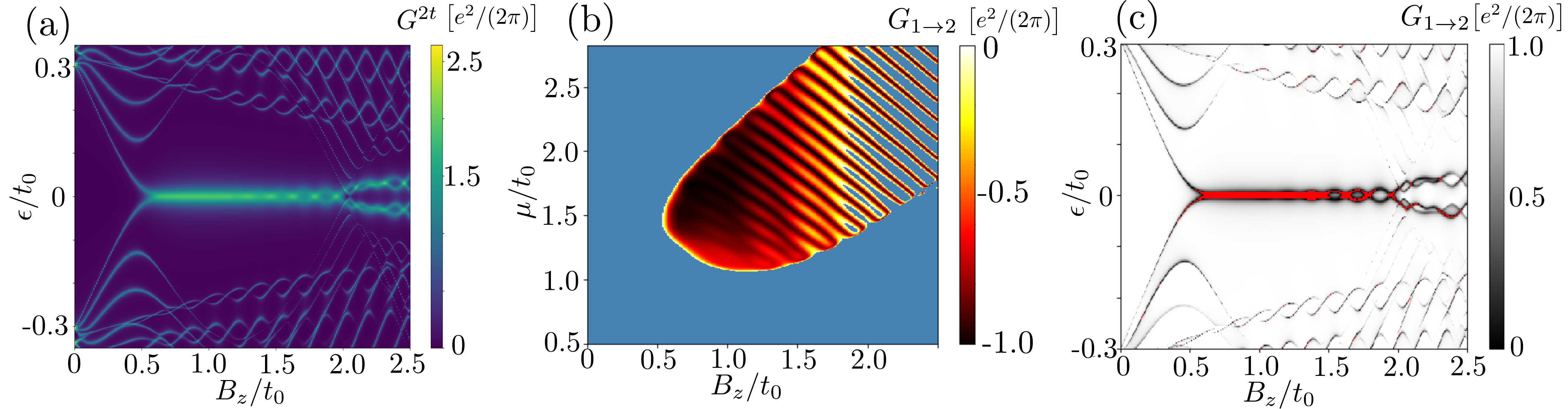}

\label{Fig:cond}

\caption{\textbf{Transport measurements.} (a) Two-terminal conductance as function of energy $\epsilon$ and Zeeman field $B_z$. (b-c) Multi-terminal conductance between contacts $1$ and $2$ with respect to Fig.~\ref{Fig:1} (a), as a function of $\mu$ and $B_z$ (b), $\epsilon$ and $B_z$ (c), respectively. In (b), all values $G_{1\rightarrow 2}>0$ are colored in blue. In (c), all values $G_{1\rightarrow 2}<0$ are colored in red. Other parameters of the plots are: $L=20 \hbar v_F/t_0$, $\Delta/t_0=0.3$, $\mu/t_0=\sqrt{2}$ ((a) and (c)), $\epsilon = 0$ (b), $t_0=t_c=1$ $v_F=1$. For computational reasons, the delta distribution separating the anti-wire from the leads is replaced with its step function approximation $\delta_{a}(x) = \mathrm{rect}(x/a)/a$ with $a=0.1$. Moreover, $T=1.5$ for (a-b) and $T=2$ for (c).}

\end{figure*}
\section{Transport characterization}
\label{sec:transport}
Since the Majorana modes are naturally embedded into a two-dimensional topological insulator, it is straightforward to bring them in proximity to other boundaries of the sample. In particular, as shown in Fig.~\ref{Fig:1} (a), it is possible to develop a weak tunnel coupling between the ends of the anti-wire and gapless helical edges. The latter, which feature up to micrometer-size mean free paths in high-quality HgTe-based QSH systems \cite{Bendias2018}, can be used as probes to perform particular transport measurements, taking advantage of their helical nature.  
In order to study the transport, we consider the amplitude $T$ in Eq.~(\ref{Eq:HT}) to be finite. The Hamiltonian of the whole system (i.e. anti-wire and helical probes) thus reads
\begin{equation}
H_{\mathrm{set}}=\int_{-\infty}^{+\infty} dx [\mathcal{H}_p+\mathcal{H}_T ] + \int_{0}^{L}dx\, [\mathcal{H}_{t_0}+\mathcal{H}_{t_c}+\mathcal{H}_{\Delta}+\mathcal{H}_{B}],
\end{equation}
where the kinetic terms for $x<0$ and $x>L$ describe the two outer helical edges.

We discuss two distinct transport schemes. The first one aims at obtaining the two-terminal conductance. In this scenario, contact $1$ and $2$ ($3$ and $4$) of Fig.~\ref{Fig:1} (a) are treated as one lead, say $1\! 2$ ($34$). Then, we have
\begin{equation}
G^{2t} = \frac{\mathrm{d}I_{1\!2}}{\mathrm{d}V_{1\!2}},
\end{equation}
where $I_{1\!2}$ is the current exiting terminals $1$ and $2$ [see Fig.~\ref{Fig:1} (a)] and $V_{1\!2}$ is their common bias with respect to the grounded superconductor. In this scheme, the two terminals and the helical edge connecting them thus act as a single tunneling probe. For small bias, we calculate $G^{2t}$ in terms of elements of the corresponding scattering matrix \cite{Fisher1981}
\begin{equation}
\label{Eq:cond1}
G^{2t}=\frac{e^2}{2\pi}\left[2+\!\sum_{j\in 1,2} \big[\vert r^{\mathrm{eh}}_{1\!2,j}\vert^2 - \vert r^{\mathrm{ee}}_{1\!2,j}\vert^2 \big]\right],
\end{equation}
where $r_{1\!2,j}^{\mathrm{e}\nu}$ are normal ($\nu = \mathrm{e}$) and  Andreev reflection amplitudes  ($\nu = \mathrm{h}$) in lead $1\!2$ in edge $j$. The elements of the scattering matrix are computed by integration of $H_{\mathrm{set}}$. Fig.~\ref{Fig:cond} (a) shows the two-terminal conductance $G^{2t}$ as a function of excitation energy $\epsilon$ 
and applied Zeeman field $B_z$. Whenever an anti-wire bound state is on resonance, a peak in the two-terminal conductance emerges. As expected, the Majorana clearly manifests itself with a strong zero-energy peak, whose properties have been extensively studied in the literature. Importantly, such a signature is not exclusively associated with the presence of Majoranas and it is thus not sufficient as a proof for their existence  \cite{SanJose2016, DasSarma2017, Stanescu2018, Fleckenstein2018Majorana, Cayao2019,DasSarma2020a,DasSarma2020b}.

In order to go beyond the simple zero-bias peak, we devise a different transport scheme which exploits the helical nature of our tunneling probe. In particular, we consider the multi-terminal conductance between contacts $1$ and $2$ [see Fig.~\ref{Fig:1} (a)] which reads 
\begin{eqnarray}
\label{Eq:G12}
G_{1\rightarrow 2}=\frac{\mathrm{d}I_2}{\mathrm{d}V_1}=\frac{e^2}{2\pi}\left[\vert t^{\mathrm{ee}}_2\vert^2-\vert t^{\mathrm{eh}}_2\vert^2\right].
\end{eqnarray}
Importantly, $G_{1\rightarrow 2}$ can either take positive or negative values, depending on which scattering process dominates: electron tunneling or crossed Andreev reflection. In the following, we demonstrate that a negative signal at zero energy can be unambiguously associated with the presence of a Majorana bound state. This statement is supported by Figs. \ref{Fig:cond} (b-c) which show that, when the anti-wire is in the topological phase and features Majoranas at its ends, the multi-terminal conductance $G_{1\rightarrow 2}$ at zero-energy is indeed negative.
Moreover, Fig.~\ref{Fig:cond} (c) shows that the negative signal (highlighted in red) is prominently seen at zero energy. There are, however, also isolated scattering events at non-zero energy with the same property. To better understand which additional information about the system can be deduced from the multi-terminal conductance, with respect to two-terminal transport, we investigate a simpler (toy) model which still describes the essential physics. This allows us to properly clarify the meaning of a negative multi-terminal conductance.

\section{Negative multi-terminal conductance and the existence of Majorana modes}
\label{sec:toy_model}
Our goal is two-fold: (i) We want to prove that the presence of a Majorana scatterer always leads to a negative multi-terminal conductance $G_{1\rightarrow 2}$. (ii) We want to clarify under which circumstances the measurement of a negative $G_{1\rightarrow 2}$ represents an unambiguous signature of the existence of a Majorana mode. 

\begin{figure}[t]
	\centering
	\includegraphics[scale=0.4]{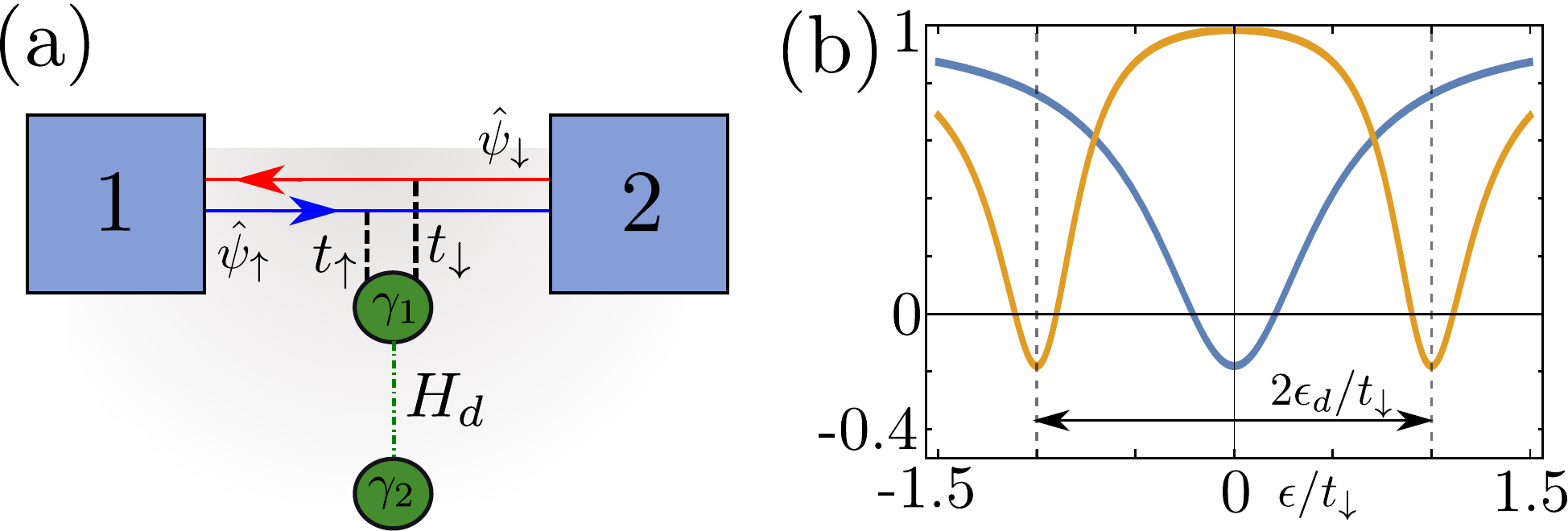}
	\label{Fig:toy_model}
	\caption{\textbf{Majorana scatterer on the helical edge}. (a) Schematic illustration of a Majorana mode $\gamma_1$ side-coupled to a helical edge. (b) Multi-terminal conductance as a function of energy $\epsilon$ with $v_F=1$, $t_{\uparrow}=1.2 t_{\downarrow}$ and $t_{\downarrow}=0.2$ and $\epsilon_d/t_{\downarrow} =0,1$ (blue, orange).
	}
\end{figure}

We consider the simple model sketched in Fig.~\ref{Fig:toy_model}(a). It consists of a single helical edge described by the Hamiltonian density
\begin{equation}
	\mathcal{H}_p^{(\nu=1)} = \sum_{\sigma}\hat{\psi}_{1,\sigma}^{\dagger}(x) (-iv_F\sigma\partial_x-\mu) \hat{\psi}_{1,\sigma}(x),
\end{equation}
which connects the leads $1$ and $2$. At $x=0$, it is tunnel coupled with a single Majorana scatterer $\hat \gamma_1 = \hat d + \hat d^\dagger$ 
via
\begin{equation}
\label{eq:Hc}
H_c=\sum_{\sigma}t_{\sigma}\big[\hat{\gamma}_1\hat{\psi}_{1\sigma}(0)+\mathrm{h.c.}\big]. 
\end{equation}
The spin-dependent coupling constants $t_\sigma$ accounts for the spin-texture of the Majorana mode \cite{AguadoMajoranaSpin2017, LossMajoranaSpin2018}. We consider a second Majorana mode $\hat{\gamma}_2 = i\hat d-i \hat d^\dagger$ which is not directly coupled to the helical edge but can (weakly) hybridize with $\hat{\gamma}_1$ via $H_{d}=-i\epsilon_d\hat{\gamma}_1\hat{\gamma}_2$. To determine the transport properties according to Eq.~(\ref{Eq:G12}), we need to compute the scattering matrix of the system \cite{Affleck2016} (see App. \ref{sec:app_scattering_matrix}). We obtain the analytical results

\begin{align}
\label{Eq:transmission_elements}
t_2^{eh} &= - \frac{t_{\uparrow}^2}{t_{\uparrow}^2+t_{\downarrow}^2\!+\!iv_F(\epsilon_d^2-\epsilon^2)/\epsilon},\\
\label{Eq:transmission_elements_B}
t_2^{ee} &=-1-t^{eh}_2
\end{align}
where $\epsilon$ is the energy at which the scattering process takes place. For $\epsilon$ sufficiently close to $\pm\epsilon_d$, we find that $t_\uparrow>t_\downarrow$ implies $G_{1\rightarrow 2}<0$. By contrast, we can show that $t_\uparrow<t_\downarrow$ leads to $G_{1\rightarrow 2}>0$ but $G_{2\rightarrow 1}<0$ (see Apps. \ref{sec:app_scattering_matrix} and \ref{sec:app_validity}). Hence, as long as the Majorana has a spin texture which is not polarized perpendicular to the spin quantization axis $z$, one of the two multi-terminal conductances $G_{1\rightarrow 2}$ or $G_{2\rightarrow 1}$ have to be negative. 

This is confirmed by Fig.~\ref{Fig:toy_model} (b), which shows $G_{1\rightarrow 2}$ for $t_{\uparrow}=1.2 t_{\downarrow}$. Without hybridization (blue line) the negative signal is centered around the Majorana energy $\epsilon=\epsilon_d=0$. The width of the dip is controlled by the magnitude of the coupling constant. Even in presence of a finite hybridization energy $\epsilon_d>0$ (orange line), the negative conductance is still present and centered around $\epsilon = \pm \epsilon_d$. Importantly, we observe that in the anti-wire, the interplay between the competing Zeeman field $\mathcal{H}_B$ and the spin-flipping scattering $\mathcal{H}_{t_c}$ guarantees that the Majoranas do not feature a spin-texture perpendicular to the $z$-axis. Therefore, we conclude that the presence of an isolated Majorana in the anti-wire necessarily leads to a negative multi-terminal conductance. 

\begin{figure}[t]
	\centering
	\includegraphics[scale=0.4]{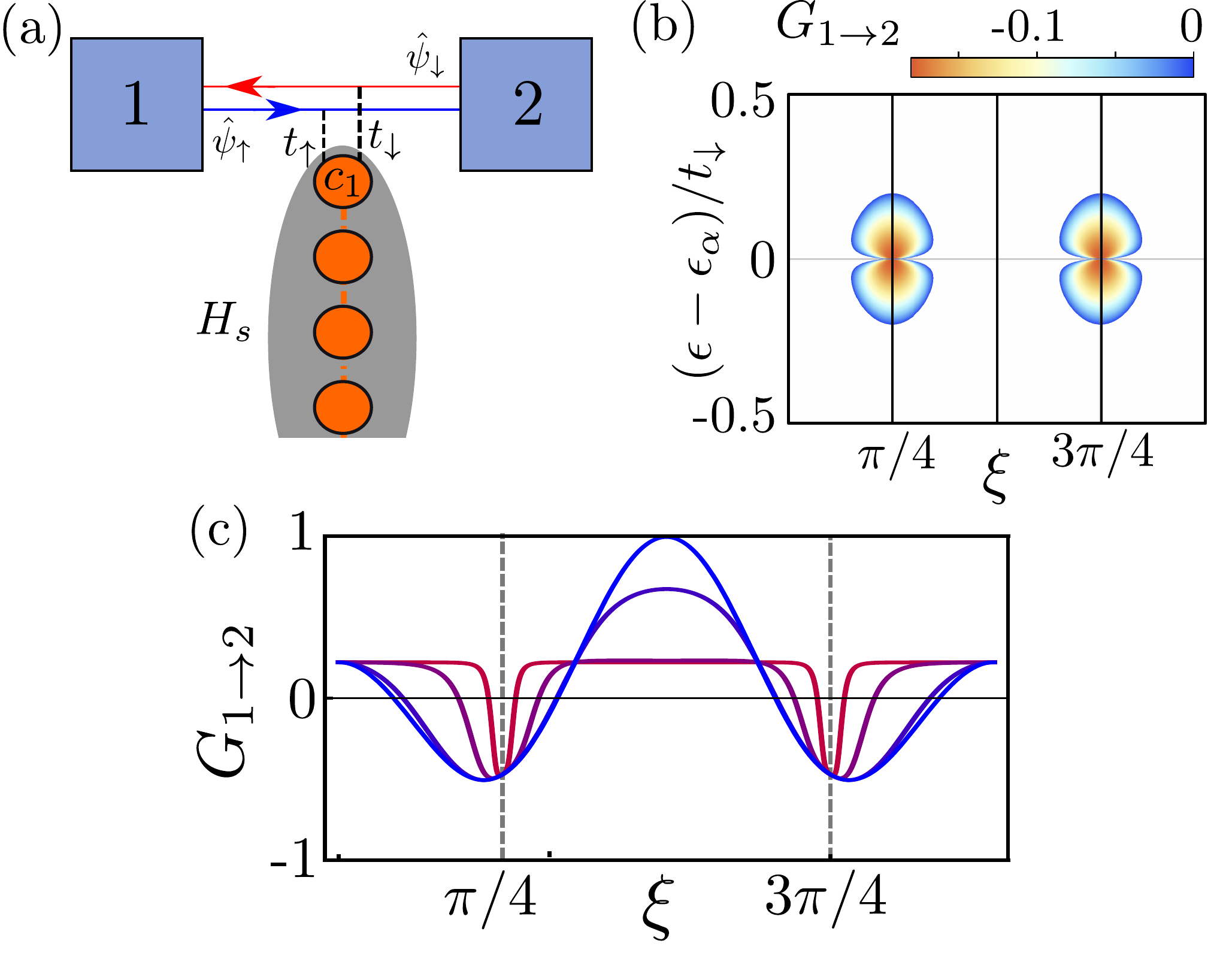}
	\label{Fig:toy_model2}
	\caption{ \textbf{Generic scatterer on the helical edge}. (a) Schematic illustration of the coupling between the helical edge and the particle-hole-symmetric system $S$. (b) Conductance $G_{1\rightarrow 2}$ for $\epsilon_{\alpha}=0$, as a function of $\xi$ and $\epsilon$. Only negative values of $G_{1\rightarrow 2}$ are shown. The parameters are $t_{\uparrow}=1.2 t_{\downarrow}$ with $t_{\downarrow}=0.2$. (c) Multi-terminal conductance $G_{1\rightarrow 2}$ for the coupling to a generic eigenstate of a PHS system on resonance $\epsilon=\epsilon_{\alpha}$, as a function of $\xi$. The different lines correspond to  $\epsilon_{\alpha}/t_{\uparrow}=2,~0.2,~0.02,$ and $0.002$ (blue to red). Further parameters are $t_{\downarrow}=3/5 t_{\uparrow}$ with $t_{\uparrow}=0.5$.}
\end{figure}

We now discuss the opposite implication, eventually showing that a negative signal at zero-energy represents an unambiguous signature of a Majorana mode. To this end, we need to consider the coupling of the helical edge with a more general particle-hole-symmetric system $S$. The latter, described by the Hamiltonian $H_S$, features several single-particle eigenstate $|\zeta_j\rangle$ with energy $\epsilon_j$. As sketched in Fig.\ \ref{Fig:toy_model2}(a), we consider the point-like tunneling at $x=0$ between the edge and a specific fermionic site of $S$, which we denote $c_1$. If we restrict our attention to a specific energy level $\epsilon_{\alpha}$, its effect on the multi-terminal conductance can be computed by considering the effective system Hamiltonian $H_S^{(\alpha)} = \epsilon_\alpha d^\dagger_\alpha d_\alpha$ and the effective tunneling Hamiltonian
\begin{equation}
\label{eq:Ht}
H_t = \sum_\sigma t_\sigma \left[ (\zeta_{\alpha,1}^{(e)*} d_\alpha^\dagger + \zeta_{\alpha,1}^{{(h)}} d_\alpha ) \psi_{1\sigma}(0) + \text{h.c.}\right],
\end{equation}
where the coefficient $\zeta_{\alpha,1}^{(e)}$ ($\zeta_{\alpha,1}^{(h)}$) represents the particle (hole) component of the state $|\zeta_\alpha\rangle$ on site $c_1$.  As before, the spin-dependent tunneling amplitudes $t_\sigma$ effectively take into account the (possible) spin-texture of the state $|\zeta_\alpha\rangle$. A careful demonstration of the validity of Eq.\ \eqref{eq:Ht} is provided in App. \ref{sec:app_pwave}, where we explicitly consider the system $S$ as a Kitaev chain. We parametrize
\begin{align}
\zeta_{\alpha,1}^{(e)} &= \Upsilon_1 \cos(\xi)\\
\zeta_{\alpha,1}^{(h)} &= \Upsilon_1 \sin(\xi),
\end{align}
neglecting a possible complex phase which has no effect on the results. The parameter $\Upsilon_1$, characterizes which fraction of the eigenstate $|\zeta_\alpha\rangle$ is localized on the site $c_1$ and its only effect is to renormalize the coupling constants. As for $\xi$, it controls whether such a fraction is more electron- or hole-like. In particular, for $\xi=0$, $H_t$ describes the coupling of the helical edge with an electronic state while, for $\xi=\pi/4$, it describes the coupling with the Majorana considered in Eq.\ \eqref{eq:Hc}. 

The multi-terminal conductance $G_{1\to2}$ associated with the effective tunneling Hamiltonian $H_t$ is plotted in Fig.\ \ref{Fig:toy_model2} (b,c). Close to resonance $\epsilon \simeq \epsilon_\alpha$, the multi-terminal conductance is negative provided that $\xi$ is sufficiently close to the Majorana case, i.e. $|\xi- \pi/4 \; (\text{mod } \pi)|\leq \bar \xi$. The threshold $\bar \xi$, depends on the detuning $(\epsilon-\epsilon_\alpha)/t_\downarrow$ as well as on the energy of the eigenstate $\epsilon_{\alpha}/t_\downarrow$. In general, $\bar \xi$ is not particularly small and the multi-terminal conductance can be negative even for values of $\xi$ which significantly differ from the Majorana case. See, for example, the blue lines in Fig.\ \ref{Fig:toy_model2} (c). This justifies the presence of isolated red spots in Fig.\ \ref{Fig:cond}(c) at high energies, even when the presence of Majorana is not expected. Importantly, however, for $\epsilon_{\alpha} = \epsilon \to 0$, the threshold goes to zero $\bar \xi \to 0$. In this case, a negative multi-terminal conductance provides a unambiguous signature of the Majorana mode. 

\section{Influence of time-reversal breaking terms and robustness against backscattering}
\label{sec:robustness}

As the formation of Majorana zero modes in the anti-wire requires the presence of a Zeeman field, let us discuss its effects on the helical edges that serves as probes for transport measurements. Importantly, the extension of the Zeeman coupling $\mathcal{H}_B$ [see Eq.\eqref{Eq:HB}] to the gapless helical regions outside the anti-wire (i.e. for $x<0$ and $x>L$) does not modify the entries of the scattering matrix. In App. \ref{sec:app_scattering_matrix}, we explicitly show this for the scattering amplitudes in Eqs.~(\ref{Eq:transmission_elements}) and \eqref{Eq:transmission_elements_B}.

\begin{figure}
\centering
\includegraphics[scale=0.2]{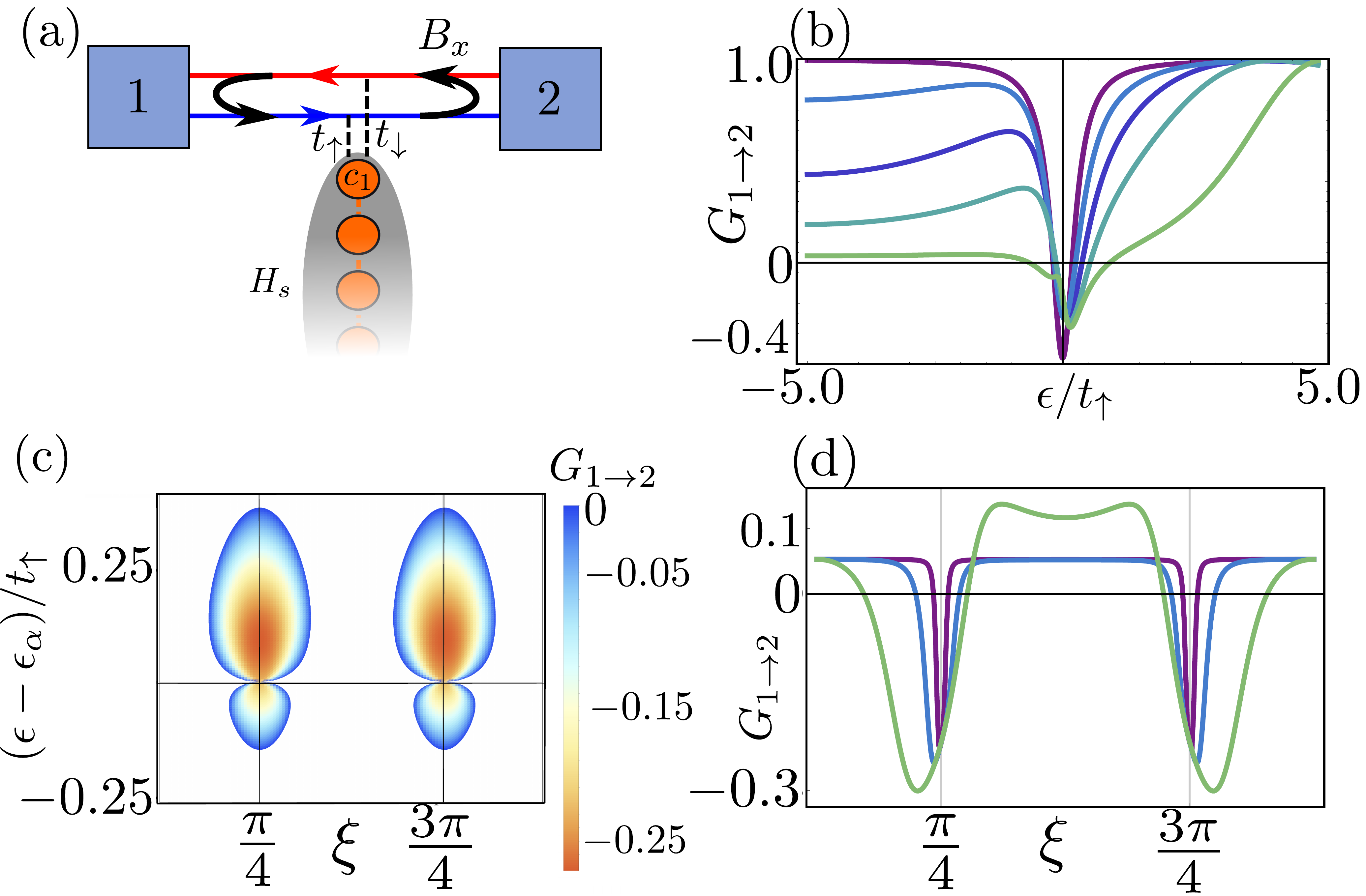}
\label{Fig:stability}
\caption{Conductance $G_{1\rightarrow 2}$ for the discussed toy-models in the presence of TR breaking backscattering: (a) Schematic of the discussed system. (b) Comparison between TR invariant (purple) and TR breaking (blue to green) transport in a helical edge, side-coupled to a Majorana. The TR breaking parameter is given by $B_x/t_\uparrow=1.25$, $ 2.5$, $3.75$, $6$ (blue to green) with $\mu/t_\uparrow=5$ and $\epsilon_d=0$. Note the for $B_x/t_\uparrow=6$, we have $B_x>\mu$ which implies that no propagating modes are present for small $\epsilon$. (c) $G_{1\rightarrow 2}$ for side-coupling a generic BdG state at energy $\epsilon_\alpha=0$ in dependence of the parametrization parameter $\xi$ and energy $\epsilon$. (d) $G_{1\rightarrow 2}$ on resonance ($\epsilon=\epsilon_\alpha$), where $\epsilon_\alpha/t_\uparrow = 5\times 10^{-2}, 5\times 10^{-3}, 5\times 10^{-4} $ (green to blue). Further parameters of the plot are: $x_i=-5 v_F/t_\uparrow$, $x_f=5 v_F/t_\uparrow$, $\mu/t_\uparrow=5$, $B_x/t_\uparrow=2.5$ ((c) and (d)), $t_\downarrow=3/5t_\uparrow$, $t_\uparrow=0.2$.}
\end{figure}

In general, however, the lack of TR symmetry spoils the topological protection of the edges and can result in the presence of backscattering, for example caused by a magnetic field along the x axis or by local impurities. This raises the question to what extent the existence of backscattering within the helical edge affects transmission and reflection amplitudes and questions the universality of the Majorana signature. To rule out possible detrimental effects due to the breaking of TR, we investigate a slightly modified version of the toy model, discussed in the latter section, where the only modification that we apply is the addition of TR breaking backscattering terms in the helical edge which is side coupled to a Majorana, a generic BdG state, respectively (see Fig.~\ref{Fig:stability} (a)).  For this model, we compute the scattering matrix and, from that, we obtain the conductance $G_{1\rightarrow 2}$ (see App. \ref{sec:app_stability}).

Fig.~\ref{Fig:stability} shows the resulting $G_{1\rightarrow 2}$ for both scenarios. Notably, finite TR breaking backscattering does not qualitatively modify the negative $G_{1\rightarrow 2}$ which represents the universal signature of a Majorana zero mode (Fig.~\ref{Fig:stability} (b)). Moreover, also for the more generic case of coupling to a general BdG state, addition of TR breaking backscattering terms does not lead to qualitative different signatures in $G_{1\rightarrow 2}$ as compared to the case without backscattering (compare Fig.~\ref{Fig:stability} (c-d) and Fig.~\ref{Fig:toy_model2} (b-c)). This is reasonable as backscattering acts in the same way to hole-like states as it does for electron-like states. In fact, any imperfection with this property is not expected to degrade the universality of the proposed signature.

\section{Discussion}
\label{sec:discussion}
\begin{figure}
	\centering
	\includegraphics[scale=1]{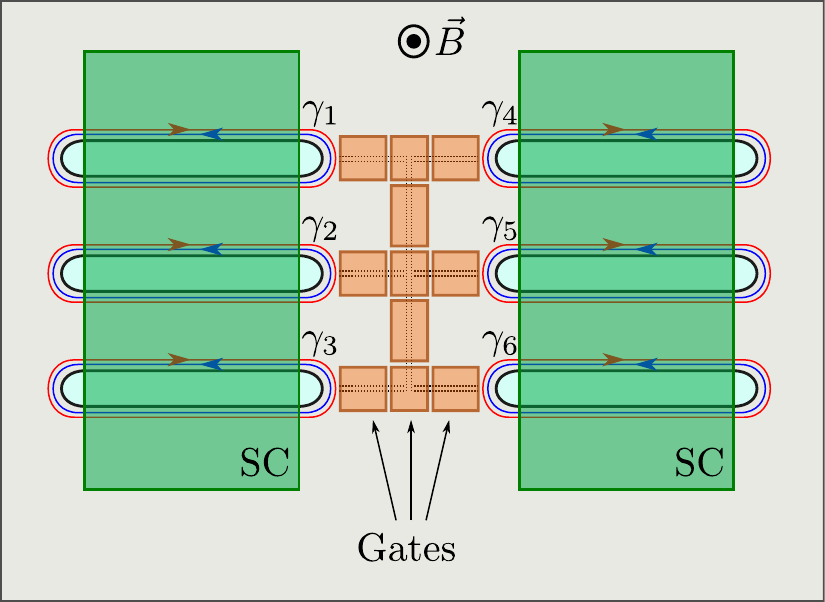}
	\label{Fig:wire_vs_antiwire}
	\caption{Coupling of six anti-wires using gate potentials
		applied to the embedding quantum spin Hall insulator (orange regions).
		}
\end{figure}

The requirements to construct isolated Majorana bound states at the helical edge, without the use of ferromagnetic barriers, are hence two pairs of helical edge modes brought into proximity with a connection in two points. As helical edge modes develop in two-dimensional topological insulators at boundaries between topological and trivial regimes, there are two ways of constructing such a system. First, cutting narrow slits in an elsewhere homogeneous two-dimensional topological insulator (Fig.~\ref{Fig:wire_vs_antiwire}). This results in what has been coined \textit{anti-wire} so far and has the advantage that, once it is possible to construct a single slit, the positioning of many slits is straightforward. Therefore, the system possesses a natural scalability, that could be of importance when it comes to quantum computations. Since different anti-wires emerge from the same underlying two-dimensional system, it is possible to tune their coupling via external gate voltages applied between two anti-wires (Fig.~\ref{Fig:wire_vs_antiwire}). Hence, the link between the two anti-wires might be changed from insulating (chemical potential inside the bulk gap of the 2D TI) to conducting (chemical potential position in conduction band), allowing for controllable fusion of the Majoranas at the end of different anti-wires. 
A second possibility to design a topologically superconducting phase is based on quantum constrictions. This setup can be obtained from the anti-wire by interchanging topological and trivial regime.

To summarize our findings, we have proposed a novel topological phase transition taking place in quantum spin Hall systems without the need of ferromagnets. This topological phase hosts topologically protected Majorana modes localized at the two ends of the anti-wire. The system we propose, being naturally hosted in a two-dimensional environment, is flexible towards scalability. Moreover, the straightforward employment of helical probes allows for more in-depth analyses of the transport properties of the system. In particular, it makes it possible to identify a novel and qualitative Majorana signature which goes beyond the standard observation of a (quantized) zero bias peak: (i) the multi-terminal conductance in the given setup carries a qualitative information based on its sign which is (ii) not expected to be influenced by particle-hole symmetric imperfections, such as backscattering processes, that are indeed detrimental for zero-bias peaks.
The experimental realization of our proposal comes with some potential challenges, in particular the realization of trenches narrow enough to allow a significant inter-edge tunneling and the coexistence of proximity-induced superconductivity with external magnetic field. However, given the recent technological developments in both directions, we believe our system to be within experimental reach.

\section*{Acknowledgements}
This work was supported by the DFG (SPP1666, SFB1170 “ToCoTronics”),
the W\"urzburg-Dresden Cluster of Excellence ct.qmat,
EXC2147, project-id 390858490, the Elitenetzwerk
Bayern Graduate School on “Topological insulators” and the Studienstiftung des Deutschen Volkes.



%

\newpage
\section*{Appendix}
\appendix

In this Appendix, we present further analysis of the calculations related to our proposal of a QSH anti-wire as a novel Majorana platform. In particular, in Sec.~\ref{sec:app_BC}, we derive the boundary conditions of the anti-wire; in Sec.~\ref{sec:app_scattering_matrix} we compute the scattering matrix of the toy model, introduced in the main text; in Sec.~\ref{sec:app_validity}, we compare the toy model with the numerical results. In Sec.~\ref{sec:app_pwave}, we justify the form of the coupling Hamiltonian used in the main text and compare our results numerically with an extended toy model on the basis of coupling to a Kitaev chain. Finally, in Sec.~\ref{sec:app_stability}, we evaluate the scattering matrix of the toy model including TR breaking terms.
\section{Derivation of the boundary conditions for the QSH anti-wire}
\label{sec:app_BC}
The kinetic Hamiltonian including impurity scattering at $x=0$ and $x=L$ can be written as
\begin{eqnarray}
\label{Eq:BC1}
\tilde{H}_{p}\!&=&  \int\mathrm{dx} \sum_{\nu,\sigma}\hat{\psi}_{\nu,\sigma}^{\dagger}(x) (-iv_F\sigma\nu\partial_x) \hat{\psi}_{\nu,\sigma}(x)\\
&+&\!T\!\int\!\mathrm{dx}\left[\delta(x)\!+\!\delta(x\!-\!L)\right]\sum_{\sigma}\left[\hat{\psi}^{\dagger}_{1,\sigma}(x)\hat{\psi}_{2,\sigma}(x)\!+\!\mathrm{h.c.}\right]\nonumber
\end{eqnarray}
with the fermionic fields $\hat{\psi}_{\nu,\sigma}(x)$ annihilating a $\nu,\sigma$ fermion at position $x$. We can formally diagonalize the Hamiltonian (\ref{Eq:BC1}) with eigenfunctions from the associated single particle problem
\begin{eqnarray}
\label{Eq:BC2}
\tilde{h}_p(x)\Psi(x)=E\Psi(x),
\end{eqnarray}
where $\tilde{h}_p(x)=-iv_F\eta_z\sigma_z\partial_x+T\left[\delta(x)+\delta(x-L)\right]\eta_x\sigma_0$ with Pauli matrices $\eta_j$, $\sigma_j$ ($j\in \{x,y,z\}$) acting on edge-, spin-space, respectively, and $\Psi(x)=(\psi_{1,\uparrow}(x),\psi_{1,\downarrow}(x),\psi_{2,\uparrow}(x),\psi_{2,\downarrow}(x))^T$. In vicinity $\delta x$ close to the impurities with $\delta x \rightarrow 0$, Eq.~(\ref{Eq:BC2}) is solved by
\begin{equation}
\label{Eq:BC3}
\Psi(\!-\delta x)=e^{\frac{T}{v_F} \eta_y\sigma_z}\Psi(\delta x),~~\Psi(L\!+\!\delta x)=e^{-\frac{T}{v_F} \eta_y\sigma_z}\Psi(L\!-\!\delta x).
\end{equation}
In the limit $T\rightarrow \infty$, this results in the boundary conditions
\begin{equation}
\label{Eq:BC4}
\begin{array}{lcr}
\psi_{1,\uparrow}(0) &=& i \psi_{2,\uparrow}(0),~~\psi_{1,\uparrow}(L)=-i\psi_{2,\uparrow}(L),\\
\psi_{2,\downarrow}(0) &=& i \psi_{1,\downarrow}(0),~~\psi_{2,\uparrow}(L)=-i\psi_{1,\downarrow}(L).
\end{array}
\end{equation}
Note that in our notation the functions $\psi_{1,\uparrow}(x)$ and $\psi_{2,\downarrow}(x)$ (and as well $\psi_{1,\downarrow}(x)$ and $\psi_{2,\uparrow}(x)$) describe states of the same chirality. Thus, we find that they obey
\begin{equation}
\begin{array}{lcr}
\psi_{1,\uparrow,q}(x) &=& i \psi_{2,\uparrow,q}(-x),\\
\psi_{2,\downarrow,q}(x) &=& i \psi_{1,\downarrow,q}(-x)
\end{array}
\end{equation}
with the plane waves $\psi_{\nu,\sigma,q}(x)=(1/\sqrt{L})\exp[i  \nu \sigma q x]$ ($\nu=(1,2)=(+,-)$ and $\sigma=(\uparrow,\downarrow)=(+,-)$) with quantized momenta $q = (\pi/L)(n-1/2)$.
By applying an expansion of the fermionic fields in terms of the functions $\psi_{\nu,\sigma,q}(x)$, namely $\hat{\psi}_{\nu,\sigma}(x)=\sum_q \psi_{\nu,\sigma,q}(x)\hat{c}_q$, we obtain the boundary condition for the fields

\begin{equation}
\label{Eq:BC5_app}
\begin{array}{lcr}
\hat{\psi}_{1,\uparrow}(x) &=& i \hat{\psi}_{2,\uparrow}(-x),\\
\hat{\psi}_{2,\downarrow}(x) &=& i \hat{\psi}_{1,\downarrow}(-x).
\end{array}
\end{equation}


Clearly, from the quantization of $q$, the fields need to be anti-periodic with respect to $2L$
\begin{eqnarray}
\hat{\psi}_{\nu,\sigma}(L)=- \hat{\psi}_{\nu,\sigma}(-L).
\end{eqnarray}
Eq.~(\ref{Eq:BC5_app}) is stated in the main as Eq.~(\ref{Eq:BC5}).


\section{Derivation of the scattering matrix}
\label{sec:app_scattering_matrix}
The system for which we aim to construct the scattering matrix is sketched in Fig.~\ref{Fig:toy_model} (a) of the main text. It is composed of three parts. The helical edge passing by the anti-wire $(\hbar = 1)$ is described by
\begin{equation}
\label{Eq:supp1}
H_p = \int\mathrm{dx} \sum_{\sigma}\hat{\psi}_{\sigma}^{\dagger}(x) (-iv_F\sigma\partial_x-\mu) \hat{\psi}_{\sigma}(x),
\end{equation}
where $\hat{\psi}_{\sigma}(x)$ are annihilating fermionic fields carrying an index $\sigma \in \{\uparrow,\downarrow\}=\{+,-\}$ and $\mu$ is a chemical potential. Since the formation of Majorana zero modes in the anti-wire requires the presence of Zeeman fields, it is a reasonable assumption to also include it in the nearby helical edge states
\begin{eqnarray}
\label{Eq:HB_sup}
H_B = \int \mathrm{dx} B_z \sum_{\sigma} \sigma\hat{\psi}^{\dagger}_{\sigma}(x)\hat{\psi}_{\sigma}(x).
\end{eqnarray}
Further, we assume a point-like coupling of the fields $\hat{\psi}_{\sigma}(x)$ to a Majorana mode $\hat{\gamma}_1$ of the anti-wire
\begin{eqnarray}
\label{Eq:supp2}
H_c=\int \mathrm{dx}~\delta(x)\hat{\gamma}_1\sum_{\sigma}t_{\sigma}\big[\hat{\psi}_{\sigma}(x)-\hat{\psi}_{\sigma}^{\dagger}(x)\big]
\end{eqnarray}
with coupling constant $t_{\sigma}$ that might depend on $\sigma$. Since TR symmetry is absent in the anti-wire, the coupling does not obey corresponding symmetry constraints. Moreover, even though hybridization of the Majoranas is exponentially suppressed in the length of the anti-wire, they might acquire a small hybridization energy
\begin{equation}
\label{Eq:supp3}
H_{d}=-i\epsilon_d\hat{\gamma}_1\hat{\gamma}_2.
\end{equation}
The two Majoranas $\hat{\gamma}_1$ and $\hat{\gamma}_2$ can be rewritten in terms of fermionic operators $\hat{d}$ and $\hat{d}^{\dagger}$ with
\begin{equation}
\label{Eq:gamma1}
\begin{array}{lcr}
\hat{\gamma}_1=\hat{d}+\hat{d}^{\dagger},\\
\hat{\gamma}_2=i\hat{d}-i\hat{d}^{\dagger}.
\end{array}
\end{equation}
Using (\ref{Eq:gamma1}), $H=H_p+H_B+H_c+H_{d}$ can also be represented as
\noindent
\onecolumngrid
\begin{eqnarray}
\label{Eq:scattering1}
H=\frac{1}{2}\!\int\!\mathrm{dx}\tilde{\Psi}^{\dagger}(x)\!
\begin{pmatrix}
-iv_F\partial_x\!-\!\mu\!+\! B_z & 0 & 0 & 0 & t_{\uparrow}(x) & t_{\uparrow}(x) \\
0 & +iv_F\partial_x\! - \!\mu\! -\! B_z & 0 & 0 & t_{\downarrow}(x) & t_{\downarrow}(x) \\
0 & 0 & -iv_F\partial_x \!+\! \mu \!-\! B_z & 0 & -t_{\uparrow}(x) & -t_{\uparrow}(x) \\
0 & 0 & 0 & +iv_F\partial_x\! +\! \mu \!+\! B_z & -t_{\downarrow}(x) & -t_{\downarrow}(x) \\
t_{\uparrow}(x) & t_{\downarrow}(x) & -t_{\uparrow}(x) & -t_{\downarrow}(x) &  \epsilon_d  & 0\\
t_{\uparrow}(x) & t_{\downarrow}(x) & -t_{\uparrow}(x) & -t_{\downarrow}(x) &  0 & -\epsilon_d \\
\end{pmatrix}\tilde{\Psi}(x)~~~~
\end{eqnarray}
\twocolumngrid
\noindent
with $\tilde{\Psi}(x)= \big(\hat{\psi}_{\uparrow}(x),\hat{\psi}_{\downarrow}(x),\hat{\psi}_{\uparrow}^{\dagger}(x),\hat{\psi}_{\downarrow}^{\dagger}(x),\hat{d},\hat{d}^{\dagger}\big)^T$
and $t_{\sigma}(x)=t_{\sigma}\delta(x)$. To diagonalize Eq.~(\ref{Eq:scattering1}), we expand $\tilde{\Psi}(x)$ in eigenfunctions of the Hamiltonian density
\begin{eqnarray}
\label{Eq:expan}
\tilde{\Psi}(x)=\sum_{k,d} U_{k,d}(x)\hat{\chi}_{k,d}
\end{eqnarray}
with matrices $U_{k,d}(x)$ and fermionic annihilation operators $\hat{\chi}_{k,d}=(\hat{C}_k,\hat{C}_d)^T$ with $\hat{C}_k = (\hat{c}_{\uparrow,k},\hat{c}_{\downarrow,k},\hat{c}_{\uparrow,k}^{\dagger},\hat{c}_{\downarrow,k}^{\dagger})$ and $\hat{C}_d=(\hat{c}_d,\hat{c}_d^{\dagger})$. Inserting Eq.~(\ref{Eq:expan}) in (\ref{Eq:scattering1}), this yields
\begin{equation}
\label{Eq:scattering11}
H=\frac{1}{2}\sum_{k,k',d,d'}\hat{\chi}_{k',d'}\int\mathrm{dx}
~U_{k',d'}^{\dagger}(x)\Xi(x)~U_{k,d}(x)\hat{\chi}_{k,d},
\end{equation}
where we defined
\begin{equation}
\Xi(x)=\begin{pmatrix}
A(x) & \eta \delta(x) \\
\eta^{\dagger}\delta(x) & \epsilon_d\sigma_z \\
\end{pmatrix}
\end{equation}
with
\begin{eqnarray}
A(x)=-iv_F\partial_x\tau_0 \sigma_z-\mu \tau_z \sigma_0 + B_z \tau_z\sigma_z
\end{eqnarray}
and
\begin{eqnarray}
\eta = \begin{pmatrix}
t_{\uparrow} & t_{\downarrow} & -t_{\uparrow} & -t_{\downarrow} \\
t_{\uparrow} & t_{\downarrow} & -t_{\uparrow} & -t_{\downarrow} \\
\end{pmatrix}^T.
\end{eqnarray}
When the columns of $U_{k,d}(x)$ are formed by orthogonal eigenfunctions of $\Xi(x)$ the problem becomes diagonal. Hence, we need to search for functions $(\Phi_k(x),\Phi_d)$, such that
\begin{equation}
\label{Eq:scattering2}
\begin{pmatrix}
A(x)\Phi_k(x) + \eta \delta(x)\Phi_d \\
\eta^{\dagger}\Phi_k(0) + \epsilon_d\sigma_z\Phi_d \\
\end{pmatrix}=\epsilon \begin{pmatrix}
\Phi_k(x)\\
\Phi_d\\
\end{pmatrix},
\end{equation}
where in the second row, we performed the integration of Eq.~(\ref{Eq:scattering1}) right away as it contains no differential forms. From Eq.~(\ref{Eq:scattering2}), we obtain an equation for the solutions $\Phi_k(x)$ by solving the second row for $\Phi_d$ and inserting the result in the first one
\begin{equation}
\label{Eq:scattering3}
A(x)\Phi_k(x)\! + \!\delta(x)\eta \begin{pmatrix}
\frac{1}{\epsilon- \epsilon_d}\! & 0 \!\\
0\! & \!\frac{1}{\epsilon+\epsilon_d} \\
\end{pmatrix} \eta^{\dagger}\Phi_k(0) =\epsilon \Phi_k(x).~~
\end{equation}
This equation might be solved in the following way \cite{Affleck2016}. When $x\neq 0$ the equation reduces to $A(x)\Phi_{k}(x)=\epsilon \Phi_k(x)$ which is solved by plane waves. Moreover, the $\delta$-distribution implies a discontinuous jump of the solutions at $x=0$. Hence, for $x>0$, $x<0$ and $x=0$, the solution takes different values. This can be incorporated by the ansatz
\begin{eqnarray}
\label{Eq:scattering4}
\Phi_k(x)=\big(\Phi^e_k(x),\Phi_k^h(x)\big)
\end{eqnarray}
with
\begin{eqnarray}
\label{Eq:scattering5a}
\Phi^{e}_k(x)&=&\begin{pmatrix}
\big(\bar{\phi}_{\uparrow}^{e}+\mathrm{sign}(x)\delta\phi_{\uparrow}^{e}\big)e^{i(k+B_z-\mu) x} \\
\big(\bar{\phi}_{\downarrow}^{e}+\mathrm{sign}(x)\delta\phi_{\downarrow}^{e}\big)e^{-i(k+B_z+\mu)x}
\end{pmatrix}\\
\Phi^{h}_k(x)&=&\begin{pmatrix}
\big(\bar{\phi}_{\uparrow}^{h}+\mathrm{sign}(x)\delta\phi_{\uparrow}^{h}\big)e^{i(k-B_z+\mu) x} \\
\big(\bar{\phi}_{\downarrow}^{h}+\mathrm{sign}(x)\delta\phi_{\downarrow}^{h}\big)e^{-i(k-B_z-\mu)x}
\end{pmatrix}
\end{eqnarray}
where
\begin{eqnarray}
\label{Eq:scattering6}
\bar{\phi}_{\uparrow/\downarrow}^{e/h}=\big(\phi_{\uparrow/\downarrow,-}^{e/h}+\phi_{\uparrow/\downarrow,+}^{e/h}\big)/2,\\
\label{Eq:scattering7}
\delta\phi_{\uparrow/\downarrow}^{e/h}=\big(\phi_{\uparrow/\downarrow,+}^{e/h}-\phi_{\uparrow/\downarrow,-}^{e/h}\big)/2.
\end{eqnarray}
Integration of Eq.~(\ref{Eq:scattering3}) using Eqs.~(\ref{Eq:scattering4}-\ref{Eq:scattering7}), this results in
\onecolumngrid
\begin{eqnarray}
\label{Eq:scattering8}
-iv_F\begin{pmatrix}
\sigma_z & 0 \\
0 & \sigma_z \\
\end{pmatrix}
\begin{pmatrix}
\phi^e_{\uparrow,+}-\phi_{\uparrow,-}^e \\
\phi^e_{\downarrow,+}-\phi_{\downarrow,-}^e \\
\phi^h_{\uparrow,+}-\phi_{\uparrow,-}^h \\
\phi^h_{\downarrow,+}-\phi_{\downarrow,-}^h \\
\end{pmatrix}
+\frac{1}{2}\eta \begin{pmatrix}
\frac{1}{\epsilon- \epsilon_d}\! & 0 \!\\
0\! & \!\frac{1}{\epsilon+\epsilon_d} \\
\end{pmatrix} \eta^{\dagger}
\begin{pmatrix}
\phi^e_{\uparrow,+}+\phi_{\uparrow,-}^e \\
\phi^e_{\downarrow,+}+\phi_{\downarrow,-}^e \\
\phi^h_{\uparrow,+}+\phi_{\uparrow,-}^h \\
\phi^h_{\downarrow,+}+\phi_{\downarrow,-}^h \\
\end{pmatrix}=0.
\end{eqnarray}
\twocolumngrid
Eq.~(\ref{Eq:scattering8}) can be reorganized such that we obtain the scattering matrix $S$
\begin{eqnarray}
\begin{pmatrix}
\phi_{\downarrow,-}^e \\
\phi_{\downarrow,-}^h \\
\phi_{\uparrow,+}^e \\
\phi_{\uparrow,+}^h \\
\end{pmatrix}
= S
\begin{pmatrix}
\phi_{\uparrow,-}^e \\
\phi_{\uparrow,-}^h \\
\phi_{\downarrow,+}^e \\
\phi_{\downarrow,+}^h \\
\end{pmatrix}
\end{eqnarray}
with
\begin{eqnarray}
\label{Eq:Smat}
S= \begin{pmatrix}
R_{--} & T_{+-} \\
T_{-+} & R_{++} \\
\end{pmatrix}
\end{eqnarray}
and
\begin{eqnarray}
R_{--}&=&\begin{pmatrix}
r_{--}^{ee} & r_{--}^{he} \\
r_{--}^{eh} & r_{--}^{hh} \\
\end{pmatrix}, ~~ R_{++}=\begin{pmatrix}
r_{++}^{ee} & r_{++}^{he} \\
r_{++}^{eh} & r_{++}^{hh} \\
\end{pmatrix},\nonumber \\
T_{+-} &=&\begin{pmatrix}
t_{+-}^{ee} & t_{+-}^{he} \\
t_{+-}^{eh} & t_{+-}^{hh} \\
\end{pmatrix}, ~~ T_{-+} = \begin{pmatrix}
t_{-+}^{ee} & t_{-+}^{he} \\
t_{-+}^{eh} & t_{-+}^{hh} \\
\end{pmatrix}.
\end{eqnarray}
For the scattering amplitudes we find
\begin{eqnarray}
R_{--}=R_{++}
\end{eqnarray}
with
\begin{eqnarray}
\label{Eq:scatteringEL1}
r_{--}^{ee}&=&r_{--}^{hh}=-r_{--}^{eh}=-r_{--}^{he}\nonumber\\
&=&\frac{t_{\uparrow}t_{\downarrow}\epsilon}{\epsilon(t_{\uparrow}^2+t_{\downarrow}^2-iv_F\epsilon)+iv_F\epsilon_d^2},\\
\label{Eq:scatteringEE}
t_{-+}^{ee}&=& t_{-+}^{hh} = \frac{t_{\uparrow}^2\epsilon}{\epsilon(t_{\uparrow}^2+t_{\downarrow}^2-iv_F\epsilon)+iv_F\epsilon_d^2}-1,\\
\label{Eq:scatteringEL}
t_{-+}^{eh} &=& t_{-+}^{he}= -\frac{t_{\uparrow}^2\epsilon}{\epsilon(t_{\uparrow}^2+t_{\downarrow}^2-iv_F\epsilon)+iv_F\epsilon_d^2},\\
t_{+-}^{ee}&=& t_{+-}^{hh} = \frac{t_{\downarrow}^2\epsilon}{\epsilon(t_{\uparrow}^2+t_{\downarrow}^2-iv_F\epsilon)+iv_F\epsilon_d^2}-1,\\
\label{Eq:scatteringEL2}
t_{+-}^{eh} &=& t_{+-}^{he}= -\frac{t_{\downarrow}^2\epsilon}{\epsilon(t_{\uparrow}^2+t_{\downarrow}^2-iv_F\epsilon)+iv_F\epsilon_d^2}.
\end{eqnarray}
With Eqs.~(\ref{Eq:scatteringEL1}-\ref{Eq:scatteringEL2}), it is easy to check that the scattering matrix of Eq.~(\ref{Eq:Smat}) is unitary. The elements of Eqs.~(\ref{Eq:scatteringEE}) and (\ref{Eq:scatteringEL}) are used in the main text. For ease of notation, in the main text, we set $T_{-+}\equiv T_{2}$ (and accordingly for its elements).

The results for the scattering amplitudes in Eqs.~(\ref{Eq:scatteringEL1}-\ref{Eq:scatteringEL2}) are independent of the values of $\mu$ and $B_z$ as both parameters do not open spectral gaps within the helical edge states passing by the anti-wire and the $\delta$-scatterer discards any dependence on the momentum of incident particles. Note that when the scatterer is modeled with a finite width $w$, for instance by replacing the $\delta$ with a Gaussian, a momentum dependence is indeed expected. Yet, this will only be significant on energy scales $v_F/w$. Thus, for small $w$ (i.e. large $v_F/w$) we expect no change in the low energy physics of our model.

In the presence of $B_z$, the symmetry protection against impurity scattering is lost as the Zeeman term breaks TR symmetry. This, however, does not influence the universality of our result as impurity scattering should affect electronic states in the same way as hole-like states. Hence, even though the transmission amplitudes might be reduced due to impurity scattering, the ratio $\vert t^{ee}_{\bar{\nu}\nu}\vert/\vert t^{eh}_{\bar{\nu}\nu}\vert$ is expected to be (on average) constant. Hence, also the multi-terminal conductance $G_{1\rightarrow 2}=\frac{e^2}{2\pi}(\vert t^{ee}_{-+}\vert^2-\vert t^{eh}_{-+}\vert^2)$, defined in the main text, is not expected to loose its qualitative information (based on its sign) in the presence of impurity scattering.
Moreover, long mean free path have been reported in the new generation of QSH systems \cite{Bendias2018}. This implies a low level of impurity scattering. We consolidate this statement more in App.~\ref{sec:app_stability}.

\section{Numerical validation of the toy model}
\label{sec:app_validity}
\begin{figure}[t]
\centering
\includegraphics[scale=0.28]{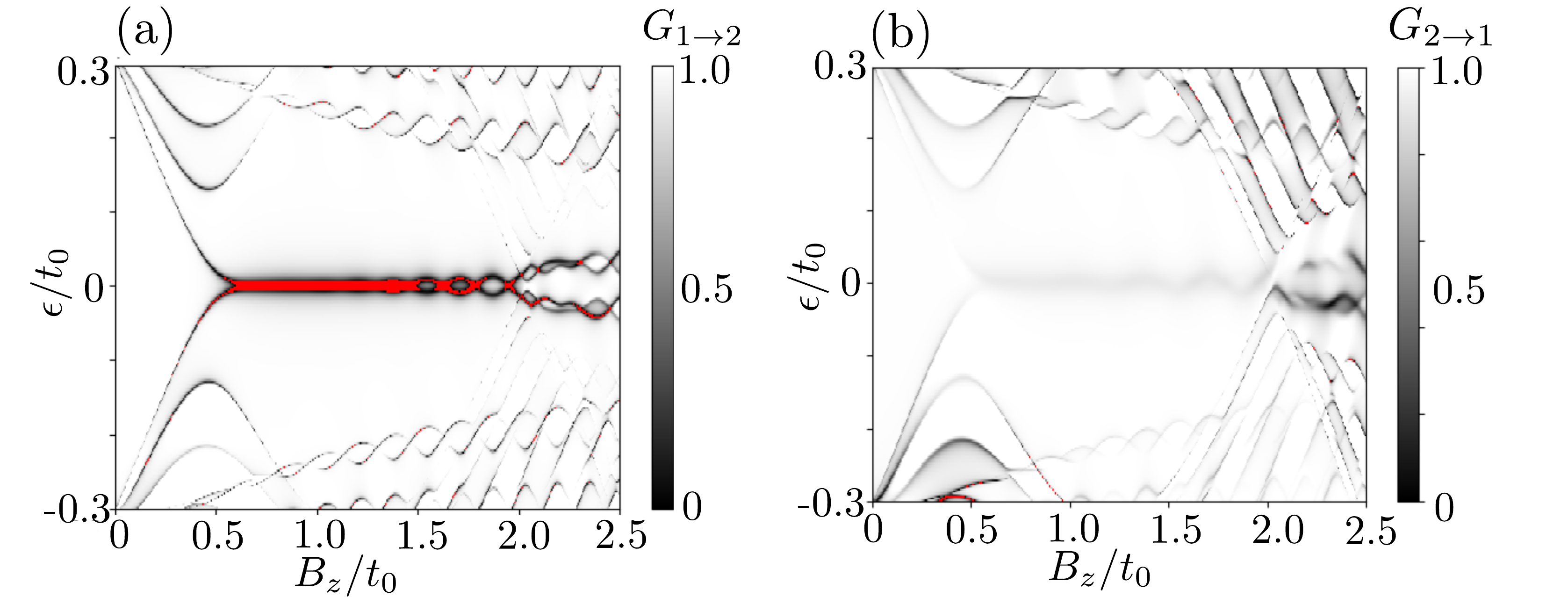}
\label{Fig:supp_cond}
\caption{Multi-terminal conductances $G_{1\rightarrow 2}$ (a) and $G_{2\rightarrow 1}$ (b) as a function of energy $\epsilon$ and $B_z$. The parameters are the same as given in Fig.~\ref{Fig:cond} of the main text. All negative values are colored in red.}
\end{figure}
As discussed in the main text, for $t_{\downarrow}> t_{\uparrow}$ in the above model, we find a multi-terminal conductance $G_{1\rightarrow 2}<0$. Likewise, the conductance $G_{2\rightarrow 1}$ is then expected to satisfy $G_{2\rightarrow 1}>0$. We can test the full model against the latter statement by numerically computing the multi-terminal conductances $G_{1\rightarrow 2}$ and $G_{2\rightarrow 1}$ using the Hamiltonian $H_{\mathrm{open}}$, defined in the main text. The results are shown in Fig.~\ref{Fig:supp_cond}. While for $G_{1\rightarrow 2}$ there is a dominant negative signal around $\epsilon=0$, for $G_{2\rightarrow 1}$ no such signal is obtained, but instead $G_{2\rightarrow 1}>0$. This confirms the validity of the employed toy model for low energies.
\section{Coupling to a p-wave superconductor}
\label{sec:app_pwave}
The toy model can also be extended for higher energies, when we do not only couple to an isolated Majorana, but to a spin-less p-wave superconductor, which, in the 1D case, can be modeled by a Kitaev chain \cite{Kitaev2001}
\begin{equation}
\label{Eq:Hd1}
H_d=\sum_{j=1}^N \mu \hat{c}_j^{\dagger}\hat{c}_j+\sum_{j=1}^{N-1} \big[(-t)\hat{c}^{\dagger}_j\hat{c}_{j+1}+\Delta\hat{c}_j^{\dagger}\hat{c}_{j+1}^{\dagger}+\mathrm{h.c.}~\big]
\end{equation}
with fermionic fields $c_j$, ($\hat{c}_j^{\dagger}$) annihilating (creating) a fermion at site $j$. The corresponding tunneling Hamiltonian can be written as
\begin{eqnarray}
\label{Eq:Hc1}
H_c =\int\mathrm{dx}\sum_{\sigma=\uparrow,\downarrow}t_{\sigma}\delta(x)\hat{\psi}_{\sigma}^{\dagger}(x)\hat{c}_1+\mathrm{h.c.},
\end{eqnarray}
where the fermions of the helical edge couple to the first site of the p-wave superconductor. Repeating the calculations of Sec.~\ref{sec:app_scattering_matrix}, with Eqs.~(\ref{Eq:Hd1}) and (\ref{Eq:Hc1}) instead of Eq.~(\ref{Eq:supp2}) and (\ref{Eq:supp3}), this results in an equation for the eigenstates of the helical edge

\begin{eqnarray}
\label{Eq:Kitaev}
\!-\!iv_F\!\begin{pmatrix}
\sigma_z & 0 \\
0 & \sigma_z \\
\end{pmatrix}\!\!
\begin{pmatrix}
\phi^e_{\uparrow,+}-\phi_{\uparrow,-}^e \\
\phi^e_{\downarrow,+}-\phi_{\downarrow,-}^e \\
\phi^h_{\uparrow,+}-\phi_{\uparrow,-}^h \\
\phi^h_{\downarrow,+}-\phi_{\downarrow,-}^h \\
\end{pmatrix}
\!+\!\frac{1}{2}\Gamma G \Gamma^{\dagger}\!\!
\begin{pmatrix}
\phi^e_{\uparrow,+}+\phi_{\uparrow,-}^e \\
\phi^e_{\downarrow,+}+\phi_{\downarrow,-}^e \\
\phi^h_{\uparrow,+}+\phi_{\uparrow,-}^h \\
\phi^h_{\downarrow,+}+\phi_{\downarrow,-}^h \\
\end{pmatrix}\!=\!0,\nonumber\\
\end{eqnarray}
where $G=[\epsilon-H_d]^{-1}$. $\Gamma$ is the Hamiltonian density of the coupling Hamiltonian $H_c$, which can be written as
\begin{equation}
H_c=\int \mathrm{dx} \delta(x)\big(
\hat{\psi}_{\uparrow}^{\dagger}(x),
\hat{\psi}_{\downarrow}^{\dagger}(x),
\hat{\psi}_{\uparrow}(x),
\hat{\psi}_{\downarrow}(x)\big) \Gamma \begin{pmatrix}
\hat{c}_1 \\
\hat{c}_1^{\dagger} \\
\vdots \\
\hat{c}_N^{\dagger}
\end{pmatrix}
\end{equation}
with
\begin{eqnarray}
\Gamma = \begin{pmatrix}
t_{\uparrow} & 0 & 0 & \dots & \dots &  0\\
t_{\downarrow} & 0 & 0 & \ddots & &  0\\
0 & -t_{\uparrow} & 0 &  &  \ddots& 0\\
0 & -t_{\downarrow} & 0 & \dots &  \dots & 0 \\
\end{pmatrix}.
\end{eqnarray}
\begin{figure}
\centering
\includegraphics[scale=0.2]{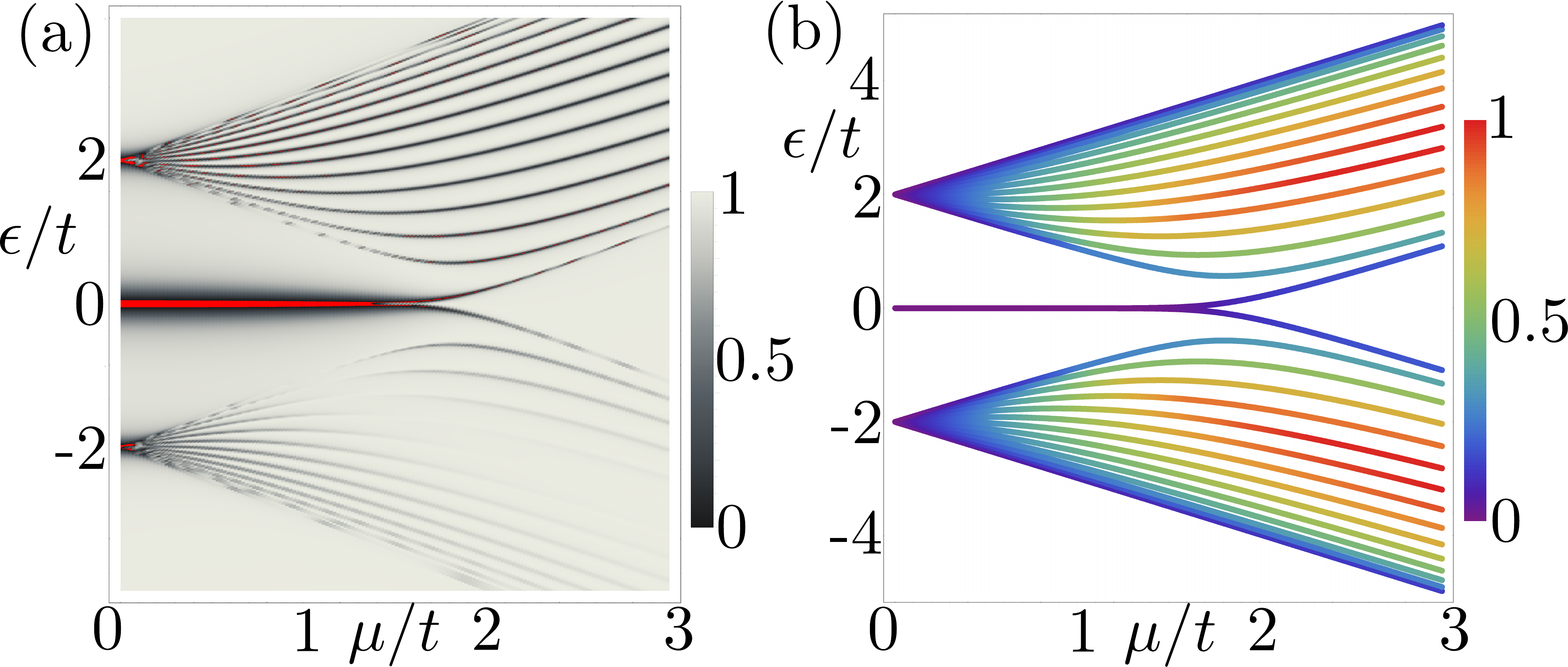}
\label{Fig:kitaev}
\caption{(a) Multi-terminal conductance $G_{1\rightarrow 2}$ for a Kitaev chain, side-coupled to a helical edge as a function of the chains chemical potential $\mu$ and energy $\epsilon$. Negative values are colored red. (b) Eigenstates of the Kitaev chain as a function of the system parameter $\mu$ and $\epsilon$. The colorcode represents the absolute difference of electronic ($\zeta_{\alpha,1}^{(e)}$) and hole-like wavefunction ($\zeta_{\alpha,1}^{(h)}$) at the first site of the chain  normalized to the maximum value reached for all eigenstates indexed by $\alpha$. Further parameters of the plots are:  $t=\Delta=0.5$, the number of sites is $N=15$.}
\end{figure}
From Eq.~(\ref{Eq:Kitaev}), we can compute the scattering matrix for the modes $\phi_{\uparrow/\downarrow,\pm}^{e/h}$, from which we obtain the conductance $G_{1\rightarrow 2}$. The results are depicted in Fig.~\ref{Fig:kitaev} (a). In accordance with the main text and the toy model of Sec.~\ref{sec:app_scattering_matrix}, we find for the topological regime $\mu<2\vert t\vert$ a prominent negative signal around $\epsilon=0$, signaling the presence of the Majorana. However, even higher energy states (in particular close to $\mu=0$) can return a negative signal.

To understand this result, we investigate again Eq.~(\ref{Eq:scattering1}), which, for the present case, takes the form
\begin{eqnarray}
\label{Eq:effective1}
H=\frac{1}{2}\int\mathrm{dx}\tilde{\Psi}^{\dagger}(x)\begin{pmatrix}
h_p & \Gamma \delta(x)\\
\Gamma^{\dagger} \delta(x) & h_d \\
\end{pmatrix}
\tilde{\Psi}(x)
\end{eqnarray}
with $h_p$ and $h_d$ the Hamiltonian density of the helical edge and the Kitaev chain and $\tilde{\Psi}(x)=\big(\hat{\psi}_{\uparrow}(x),\hat{\psi}_{\downarrow}(x),\hat{\psi}_{\uparrow}^{\dagger}(x),\hat{\psi}_{\downarrow}^{\dagger}(x), ~\hat{c}_1,\hat{c}_1^{\dagger},\dots,\hat{c}_N^{\dagger}\big)$ \cite{note_supp}. We can now apply a unitary transformation to Eq.~(\ref{Eq:effective1}) that diagonalizes $h_d$
\begin{eqnarray}
F=\begin{pmatrix}
\mathbb{1} & 0 \\
0 & U_d \\
\end{pmatrix}.
\end{eqnarray}
Then, Eq.~(\ref{Eq:effective1}) becomes
\begin{equation}
\label{Eq:effective1}
H=\frac{1}{2}\int\mathrm{dx}\tilde{\Psi}^{\dagger}(x)F\begin{pmatrix}
h_p & \Gamma U_d \delta(x) \\
U_d^{\dagger}\Gamma^{\dagger} \delta(x) &U_d^{\dagger} h_d U_d\\
\end{pmatrix}
F^{\dagger}\tilde{\Psi}(x).
\end{equation}
Since $U_d$ diagonalizes $h_d$, it is formed from the eigenstates of $h_d$
\begin{equation}
U_d = \big(\zeta_1, \zeta_2, \dots \zeta_{2N}\big),
\end{equation}
where $\zeta_{\alpha}=(\zeta_{\alpha,1}^{(e)},\zeta_{\alpha,1}^{(h)},...,\zeta_{\alpha,N}^{(e)},\zeta_{\alpha,N}^{(h)})^T$ are column vectors with the property $h_d\zeta_{\alpha}=\epsilon_{\alpha}\zeta_{\alpha}$. The transformed coupling Hamiltonian thus contains the elements of the eigenfunctions at the first site. Consequently, in a low energy approximation around an eigenenergy $\epsilon_{\alpha}$ of $h_d$, the coupling only happens to the first site of the corresponding eigenstate $\zeta_{\alpha}$. If we want to preserve particle-hole symmetry, it also has to connect to its particle-hole partner at $-\epsilon_{\alpha}$, $\hat{P}\zeta_{\alpha}$ with the particle-hole operator $\hat{P}=\mathbb{1}_{N\times N}\otimes \sigma_x\hat{K}$, where $\hat{K}$ denotes complex conjugation. The effective Hamiltonian thus reads
\begin{eqnarray}
\label{Eq:effective2}
H_{\alpha}=\frac{1}{2}\int\mathrm{dx}\tilde{\Psi}^{\dagger}_{\alpha}(x)\begin{pmatrix}
h_p & \Gamma_{\alpha} \delta(x)\\
\Gamma_{\alpha} \delta(x) &\epsilon_{\alpha}\sigma_z\\
\end{pmatrix}
\tilde{\Psi}_{\alpha}(x)
\end{eqnarray}
with the basis
$\tilde{\Psi}_{\alpha}= \big(\hat{\psi}_{\uparrow}(x),\hat{\psi}_{\downarrow}(x),\hat{\psi}_{\uparrow}^{\dagger}(x),\hat{\psi}_{\downarrow}^{\dagger}(x),\hat{d}_{\alpha}, \hat{d}_{\alpha}^{\dagger}\big)$ where $\hat{d}_{\alpha}^{\dagger}$ creates a fermion at energy $\epsilon_{\alpha}$. The coupling matrix $\Gamma_{\alpha}$ is given by
\begin{eqnarray}
\Gamma_{\alpha}\!\!=\!\! \begin{pmatrix}
t_{\uparrow} \zeta_{\alpha,1}^{(e)} & t_{\downarrow}\zeta_{\alpha,1}^{(e)} & -t_{\uparrow} \zeta_{\alpha,1}^{(h)} & -t_{\downarrow} \zeta_{\alpha,1}^{(h)} \\
t_{\uparrow} \zeta_{\alpha,1}^{(h)*} & t_{\downarrow}\zeta_{\alpha,1}^{(h)*} & -t_{\uparrow} \zeta_{\alpha,1}^{(e)*} & -t_{\downarrow} \zeta_{\alpha,1}^{(e)*} \\
\end{pmatrix}^T.
\end{eqnarray}
As discussed in the main text, this effectively corresponds to the coupling to a particle $\chi^{\dagger}=\zeta_{\alpha,1}^{(e)*}d_{\alpha}^{\dagger}+\zeta_{\alpha,1}^{(h)}\hat{d}_{\alpha}$. In particular, for $\zeta_{\alpha,1}^{(e)}\equiv \zeta_{\alpha,1}^{(h)}=1$, it corresponds to the toy model of Sec.~\ref{sec:app_scattering_matrix}. 
On the basis of the effective model of Eq.~(\ref{Eq:effective2}) we find (as discussed in the main text) two main results: (i) away from zero-energy a negative signal in the multi-terminal conductance $G_{1\rightarrow 2}$ is reached whenever the form of the particle $\chi$ deviates less than a threshold $\bar{\xi}$ from the Majorana from, i. e. whenever $\delta\zeta =\vert \vert \zeta_{\alpha,1}^{(e)}\vert-\vert\zeta_{\alpha,1}^{(h)}\vert\vert\leq \bar{\xi}$ and, more importantly, (ii) as $\epsilon_{\alpha}\rightarrow 0$ also the threshold $\bar{\xi}\rightarrow 0$.

We can numerically confirm our analysis when analyzing the situation of the side-coupled Kitaev chain. Fig.~\ref{Fig:kitaev} (b) visualizes the (numerically) obtained values of $\delta\zeta$ for each eigenstate (on the first site). At $\mu= 0$, each eigenstate of the Kiteav chain satisfies the Majorana condition at the first site. Hence, we expect to find a negative multi-terminal conductance for all eigenstates, which coincides with the numerical results in Fig.~\ref{Fig:kitaev} (a). Away from $\mu =0$, eigenstates at $\epsilon \neq 0$ successively loose the Majorana condition and the dominant negative signal in the multi-terminal conductance is as well lost for those states.
At zero-energy, however, the Majorana form is kept throughout the whole topological phase and likewise also the negative signal persists.

\section{Stability against time-reversal breaking scattering}
\label{sec:app_stability}
In App. \ref{sec:app_scattering_matrix}, we have already seen that a TR symmetry breaking Zeeman field $B_z$ does not influence the universality of the obtained conductance signature (i.e. negative $G_{1\rightarrow 2}$ in the presence of the Majorana at zero energy). This suggests that TR symmetry is not among the determinative symmetries to eventually obtain negative $G_{1\rightarrow 2}$. Yet, one may wonder if this stems from the observation that $B_z$ does not induce TR breaking backscattering. To rule out this possibility, we now discuss the influence of such backscattering terms. 

The model we analyze is given by Eq.~(\ref{Eq:scattering3}), i.e.
\begin{eqnarray}
\label{Eq:stability1}
\tilde{A}(x)\Phi_k(x)\! + \!\delta(x)\eta \begin{pmatrix}
\frac{1}{\epsilon- \epsilon_d}\! & 0 \!\\
0\! & \!\frac{1}{\epsilon+\epsilon_d} \\
\end{pmatrix} \eta^{\dagger}\Phi_k(0) =\epsilon \Phi_k(x),~~
\end{eqnarray}
where $\tilde{A}(x)=-iv_F\tau_0\sigma_z\partial_x -\mu \tau_z\sigma_0 + \tau_z\sigma_x B_x $ now contains TR breaking backscattering contributions $B_x$. 
Away from $x=0$, Eq.~(\ref{Eq:stability1}) is solved by integration

\begin{equation}
\Phi_k(x_b)\!=T_B(x_b,x_a)\Phi_k(x_a),~
\end{equation}
where 
\begin{equation}
\label{Eq:stability1a}
T_B(x_b,x_a)=\!\exp\left[\!\frac{i}{v_F}\!\int_{x_a}^{x_b}\!\mathrm{d}x~ \tau_0\sigma_z\!\left(\!\epsilon \!-\!(B_x\tau_z\sigma_x\!-\!\mu\tau_z\sigma_0)\!\right)\!\right].
\end{equation}
The $\delta$ scattering event at $x=0$ requires more care. As the eigenfunctions are not expected to always possess a pure plane-wave character, the ansatz of Eq.~(\ref{Eq:scattering4}) might no longer be valid. Still, integration from $x=-\varepsilon$ to  $x=\varepsilon$ and taking the limit $\varepsilon\rightarrow 0$ yields a defining equation for scattering at the $\delta$-barrier, given by
\begin{eqnarray}
\label{Eq:stability2}
-i v_F \tau_0\sigma_z \left[\Phi_k(0^+)-\Phi_k(0^-)\right]+\eta G_d \eta^{\dagger}\Phi_k(0)=0,~~~
\end{eqnarray} 
where we introduced the shorthand notation
\begin{eqnarray}
G_d= \begin{pmatrix}
\frac{1}{\epsilon- \epsilon_d}\! & 0 \!\\
0\! & \!\frac{1}{\epsilon+\epsilon_d} \\
\end{pmatrix}.
\end{eqnarray}
Similar to Eq.~(\ref{Eq:scattering4}), Eq.~(\ref{Eq:stability2}) can be solved with a symmetric ansatz $\Phi_k(0)=1/2\left(\Phi_k(0^+)+\Phi_k(0^-)\right)$. This automatically leads to the transfer matrix, associated with the $\delta$-barrier
\begin{eqnarray}
\Phi_k(0^+)=T_{\delta}\Phi_k(0^-),
\end{eqnarray}
where
\begin{equation}
T_{\delta}=\left[-iv_F \tau_0\sigma_z +\frac{1}{2}\eta G_d \eta^{\dagger}\right]^{-1}\left[-iv_F\tau_0\sigma_z-\frac{1}{2}\eta G_d \eta^{\dagger}\right].
\end{equation}
The transmission in a helical edge from $x_i<0$ to $x_f>0$, including backscattering by $B_x$, side-coupled to a Majorana, is then described by the compiled transfer matrix
\begin{eqnarray}
\label{Eq:stability3}
T(x_f,x_i)= T_B(x_f,0)T_\delta T_B(0,x_i).
\end{eqnarray}
From $T(x_f,x_i)$, it is straightforward to compute the associated scattering matrix and, subsequently, the conductance $G_{1\rightarrow 2}$. Moreover, it is straightforward to generalize Eq.~(\ref{Eq:stability3}) to the generic case just by replacing $\eta \rightarrow \Gamma_\alpha$.


\begin{thebibliography}{59}
\bibitem [{not()}]{note_supp}%
  \BibitemOpen
  \href@noop {} {}\bibinfo {note} {With Hamiltonian density of the Kitaev chain
  we just mean the matrix elements of Eq. (\ref{Eq:Hd1}) in the given basis
  devided by the system size.}\BibitemShut {Stop}%
\end{thebibliography}

\end{document}